\documentclass[11pt,superscriptaddress,floatfix]{article}
\usepackage{graphicx}
\usepackage{epsfig}
  \usepackage{amssymb}
 \usepackage{color}
 \usepackage{jheppub}

\begin{document}
\newcommand{\be}{\begin{eqnarray}}
\newcommand{\ee}{\end{eqnarray}}
\newcommand\del{\partial}
\newcommand\nn{\nonumber}
\newcommand{\Tr}{{\rm Tr}}
\newcommand{\Trg}{{\rm Str}}
\newcommand{\ident}{{\bf 1}}
\newcommand{\bmat}{\left ( \begin{array}{cc}}
\newcommand{\mat}{\left ( \begin{array}{cc}}
\newcommand{\emat}{\end{array} \right )}
\newcommand{\matt}{\left ( \begin{array}{ccc}}
\newcommand{\ematt}{\end{array} \right )}
\newcommand{\vect}{\left ( \begin{array}{c}}
\newcommand{\evect}{\end{array} \right )}
 \definecolor{Bittersweet}   {cmyk}{0,0.75,1,0.24}
\newcommand{\rbs}{\color{Bittersweet}}
\definecolor{Greenjv}   {cmyk}{0.3,0.0,0.5,0.5}
\newcommand{\gjv}{\color{Greenjv}}
\definecolor{Periwinkle}    {cmyk}{0.57,0.55,0,0}
\newcommand{\bl}{\color{Periwinkle}}
\newcommand{\tr}{{\rm Tr}}
\newcommand{\nb}{\bar n}
\newcommand{\eref}[1]{(\ref{#1})}

\voffset -2cm

  
\title{Bosonic Partition Functions at Nonzero (Imaginary) Chemical Potential}

\author{M. Kellerstein and J.J.M. Verbaarschot}
\affiliation{Department of Physics and Astronomy, Stony Brook University, Stony Brook
 New York 11794, USA}
\emailAdd{moshe.kellerstein@stonybrook.edu}
\emailAdd{jacobus.verbaarschot@stonybrook.edu}

\date{\today}
\abstract{
  We consider bosonic random matrix partition functions at nonzero
  chemical potential and compare the chiral condensate, the baryon number density
  and the baryon number susceptibility to the
  result of the corresponding fermionic partition function. We find
  that as long as results are finite, the phase transition of the
  fermionic theory persists in the bosonic theory. However, in case
  that bosonic partition function diverges and has to be regularized,
  the phase transition of the fermionic theory does not occur in
  the bosonic theory, and the bosonic theory is always in the broken
  phase.}

\maketitle

\newpage
\section{Introduction}

Universal random matrix behavior of QCD Dirac spectra can be understood in terms of
chiral Lagrangians and is a direct consequence of spontaneous symmetry breaking
in the presence of a mass gap so that at low energies the theory reduces to
a system of weakly interacting Goldstone modes. Spontaneous symmetry breaking
also occurs in random matrix theories in the limit of large matrices, and because
they also have  mass gap, the low energy limit of the random matrix theory partition function
reduces to an integral over ``Goldstone modes''. 
In the microscopic scaling domain,
where $ \lambda V \Sigma$ (with $\lambda$ the Dirac eigenvalue, $V$ the space-time volume and $\Sigma$ the chiral condensate) is kept fixed in the thermodynamic limit, the generating
function for
Dirac spectra of  QCD or QCD-like
theories coincides with the one obtained from random matrix theories
with the same global symmetries
and is identical to the one obtained from the corresponding chiral Lagrangian.
The reason is that, in all cases we know of, the global symmetries in QCD are broken
spontaneously in the same way as in the corresponding random matrix theory.

It has been well established that lattice QCD Dirac spectra fluctuate according
to the corresponding random matrix theory in the microscopic domain
(see \cite{tilo,ency,gernot}). Because this
agreement is based on the spontaneous breaking of the flavor symmetry, one would
expect that, as a consequence of the Coleman-Mermin-Wagner theorem, the agreement
with Random Matrix Theory in two dimensions is structurally different from
the agreement found in four dimensions. Yet this is not the case
\cite{graz,bieten1,bieten2,giusti,2d}. The picture that emerges from the two-flavor massless Schwinger model \cite{graz,bieten1,bieten2,dam-schwing},
is that the low-lying eigenvalues are correlated according to chiral Random Matrix
Theory while the chiral condensate defined in the usual way vanishes. For
two-dimensional QCD \cite{giusti}, a nonzero chiral condensate was found for U($N_c$) theories, while  for SU($N_c$) theories the mass dependence of the chiral condensate
is consistent with $ m^{(N_f-1)/(N_f+2)} $, the same as for the Schwinger model. Since
$\Pi_1(U(N_c)) = {\mathcal Z}$, the former observation could be interpreted 
in terms of a  Kosterlitz-Thouless phase.
  We performed  quenched lattice simulations of two-dimensional QCD at strong coupling \cite{2d} and found
  that the agreement of QCD Dirac spectra with random matrix theory is as good
  as  in four dimensions for comparable statistics.

  The resolvent of the Dirac operator $D$ for $N_f$ quarks with mass $m$
  can be expressed in terms of the generating function $Z(m,z,z')$ as
  \be
  G(m,z) = \left . \frac d{dz}\right|_{z'=z} Z(m,z,z')
  \ee
  with
\be
Z(m,z,z') = \left \langle \frac{ {\det}^{N_f} (D+m) \det(D+z) }{\det (D+z')}
\right \rangle .
\label{zgen}
\ee
Because of the inverse determinant  
this generating function has a noncompact
symmetry \cite{sener1}. 
  It has been argued that the Mermin-Wagner-Coleman theorem can be violated for
  noncompact continuous symmetries \cite{ziegler,zirn,zirn-spencer,seiler}. In particular,
  it has been shown that the SO(2,1) symmetry of a hyperbolic spin chain
  is spontaneously broken also in one and two dimensions. In essence, the reason
  is that a partition function with a noncompact symmetry can only be defined
  if this symmetry is spontaneously broken to its compact subgroup SO(2). In a conformal invariant
  theory the spectral density of the Dirac operator also scales as $\rho(\lambda)\sim V\lambda^\alpha$ and this scenario might reconcile conformal behavior with universal random
  matrix statistics \cite{kuti,anna,kuti2}.

  As is the case for the hyperbolic spin chain we could have the scenario that the
  compact symmetry remains unbroken, so that we have a vanishing chiral condensate, while
  the noncompact symmetry is spontaneously broken resulting in universal random matrix
  behavior. It is important to note that the chiral condensate is obtained at
  fixed $m$ in the thermodynamic limit,
while  random matrix behavior takes place on the scale of the average level spacing. Since
  the Mermin-Wagner-Coleman theorem requires  a vanishing chiral condensate 
  in two dimensions or less, we could satisfy the Banks-Casher relation if the
  low-lying eigenvalues scale as $1/V^{1/(\alpha +1)}$ with $\alpha > 0$. At the same
  time the noncompact chiral symmetry of the generating function could be
  broken spontaneously by these eigenvalues.
  
  Let us discuss what has been found in lattice simulations of the
  massless $N_f$-flavor Schwinger model.
The average macroscopic spectral density is given by $\rho(\lambda) \sim V \lambda^\alpha $ with
$\alpha =(N_f -1)/(N_f+2)$ \cite{smilga,smilga-jv}. This results in a chiral condensate that
vanishes as $m^\alpha$ for $ m\to 0$.
What transpires from  lattice simulations \cite{graz,bieten1,bieten2} is that the
  chiral condensate vanishes as predicted  while the  rescaled low-lying Dirac eigenvalues,
  $\lambda_k V^{1/(\alpha +1)}$ fluctuate according to random matrix theory. The
  low-lying eigenvalues spontaneously break the symmetry of the generating
  function but because, they scale as $1/V^{1/(\alpha +1)}$ with the volume, the
  chiral condensate remains zero. The generating function for the
  resolvent that reflects this behavior of the low-lying Dirac spectrum 
is of the form
  \be
  Z(m,z,z')= \int_{U\in G/H} dU e^{-V^{1/(\alpha+1)} \Tr M(U+U^{-1})}
\label{zeff}
  \ee
  with $M ={\rm diag}(m,\cdots,m,z,z')$ and $G\to H$ the spontaneous symmetry breaking pattern
  of the generating function.
  
  There are other possible explanations of the lattice data. For example, the
  states might be localized with a localization length that is much larger
  than the size of the system so that the eigenvalues obey random matrix
  statistics, but the chiral condensate vanishes in the thermodynamic limit.
To distinguish such scenario from the
partition function (\ref{zeff})  will require lattice simulations on very
large volumes which may not yet be feasible at this time.
 
  In this paper we study a much simpler question, namely to what extent
  spontaneous symmetry breaking in fermionic random matrix partition functions
  (averages of determinants) differs from spontaneous symmetry breaking
  in bosonic random matrix partition functions (averages of inverse determinants). This question was first studied in the context of the validity of the replica
  trick for the Gaussian Unitary Ensemble where it was shown that the partition function
  for  fermionic replicas 
  is structurally different from the partition function
  for  bosonic replica and result in a different replica limit
  \cite{critique}.
 Later this was explained in terms of the Toda lattice
  equation  which gives a two-step recursion relation in the
  number of replicas that connects bosonic and fermionic partition
  functions \cite{split-fact}.

The relation between bosonic and fermionic partition functions was also studied
  in \cite{kim-bos} for the phase quenched
  partition function.  As will be explained in section III, in the bosonic case, the 
    pion condensate is nonvanishing
    for all values of the chemical potential with a spontaneously broken
    noncompact symmetry, while in the fermionic case pions
    only condense for $\mu> m_\pi/2$.
    This section is preceded by an introduction
    of the random matrix models that will be
  studied in this paper. The one flavor partition function at
  imaginary chemical potential will be analyzed in section IV, and  we reduce the one-flavor
  bosonic partition function to a one-dimensional integral that can easily be evaluated
  numerically. In section V, we work out the one flavor bosonic partition function
  for real chemical potential at zero quark mass 
  and compare its properties to the fermionic partition
  function with the same parameters. Concluding remarks are made in section VI.
  Additional technical details are worked out in three appendices.

  A preliminary account of some aspects of the issues discussed in this paper was published
  as a contribution to Conference Proceedings \cite{moshe}.

  

\section{Random Matrix Theories}
We consider two different random matrix theories for QCD at nonzero chemical potential,
\be
D_1 &=& \bmat m\ident & C +\mu \ident \\ -C^\dagger +\mu\ident  & m\ident \emat, \qquad
\label{D1} \\
D_2 &=& \bmat m\ident & C +\mu D \\ -C^\dagger +\mu D^\dagger& m\ident \emat \label{D2}
\ee
with complex $n\times n$ matrices $C$ and $D$ distributed according to
\be
P(C) = e^{-n \Sigma^2 \Tr C C^\dagger}.
\ee
The ensemble $D_1$ was introduced in \cite{andy} for
imaginary chemical
potential  and  in \cite{misha} for real chemical potential, while the ensemble
$D_2$ was introduced in \cite{osborn}. For each of the ensembles we consider
the bosonic and fermionic one-flavor and two-flavor phase-quenched partition
functions,
\be
Z_{N_f=1} = \langle {\cal N} \det (D+m) \rangle, \label{z1} \\
Z_{N_f=-1} = \left \langle \frac 1{{\cal N}\det (D+m)} \right \rangle, \label{zm1} \\
Z_{1+1^*} = \langle |{\cal N}\det (D+m)|^2 \rangle, \label{zpq} \\
Z_{0/1+1^*} = {\cal N}\left \langle \frac 1{|{\cal N}\det (D+m)|^2}  \right\rangle, \label{zmpq}
\ee
The normalization factor is chosen such that the free energy is $\mu$ independent for small $\mu$ and $m \to 0$. It turns out that this factor is given by
\be
{\cal N} = e^{n\Sigma^2 \mu^2}.
\ee
In the microscopic domain, $m \sim 1/V$ and $\mu^2 \sim 1/V$, the mass and
chemical potential dependence of the partition functions is universal and
coincides with that of the QCD partition function. In this limit, the random matrix ensembles
$D_1$ and $D_2$ give the same results which can also be derived from
the corresponding chiral Lagrangian.
In particular, the one-flavor
partition function does not depend on the chemical potential in this domain.
Since the
chemical potential of the phase quenched fermionic partition function can be
interpreted as an isospin
chemical potential \cite{wilczeck,SS} this partition function is $\mu$-independent only up
to $\mu = m_\pi/2$ at which point a phase transition to 
 a pion condensation phase occurs. The phase quenched bosonic partition function does not have a phase transition
as a function of $\mu$ \cite{kim-bos} as will be discussed in more detail in the next section.
An imaginary chemical potential does not change the hermiticity properties of the
Dirac operator and in the microscopic domain the partition function does not depend on
it.
 
The ensemble $D_2$ does not have any other phase transitions in the nonuniversal
domain. On the other
hand, the ensemble $D_1$ has nonuniversal phase transition. For $\mu=i\mu_i$ purely
imaginary it has a second order phase transition to a chirally restored phase
at $\mu_i=1$ \cite{andy}, whereas for real $\mu$ it has a first order transition
at $\mu = 0.527$ \cite{misha}. This phase transition resembles the QCD phase transition
to a phase of nonzero baryon density which is why this model is particularly interesting.
One of the main questions of this paper is the
fate of this phase transition for the bosonic partition function.

The random matrix partition functions of both ensembles
can be evaluated by a variety of methods
such as the supersymmetric method, the replica trick, resolvent expansion
technique, the Toda lattice equation, chiral Lagrangians etc. .
 However, only the partition functions of the 
of the two-matrix ensemble $D_2$ can be evaluated  analytically at finite $n$
using orthogonal polynomial methods \cite{osborn,AHOV,kim-bos,imagmu}.
The fermionic as well as phase quenched partition functions of the ensemble
$D_1$ have been evaluated both for real \cite{misha,HJV} and imaginary
chemical potential \cite{andy,lehner}. Both exact results in terms of
one-dimensional integrals \cite{HJV} and mean field results \cite{misha,janik}
have been obtained. The bosonic partition function of the ensemble $D_1$ has
not been studied in the literature, and we will evaluate it both for
imaginary chemical potential at nonzero quark mass and real chemical
potential at zero quark mass.

\section{Phase Quenched QCD}

The phase quenched fermionic partition function can be rewritten as
\be
Z_{1+1^*} &=& \langle |{\cal N}\det (D+\mu\gamma_0+m)|^2 \rangle, \nn\label{zpq2} \\
&=&\langle |{\cal N}\det (D+\mu\gamma_0+m)\det (D-\mu\gamma_0+m) \rangle,
\ee
and is therefore the two-flavor partition function at nonzero isospin chemical
potential \cite{wilczeck}.
It has a phase transition to a Bose condensed phase at $\mu =m_\pi/2$. This transition
coincides with the point where the quark mass enters the cloud of eigenvalues 
\cite{toublan}.

The phase quenched bosonic partition function \eref{zmpq} can be evaluated simply
 by writing it
as an integral over the joint probability distribution \cite{kim-bos}
 \be
Z_{0/1+1^*}(z,z^*,\mu) = \int \prod_{k=1}^n\frac{ d^2z_k  w(z_k,z_k^*; \mu)}{(z^2-z_k^2)(z^{*\, 2}-z_k^{*\,2})}
|\Delta(z_k^2)|^2 ,
\ee
where \cite{osborn}
\be
w(z_k,z_k^*; \mu)= |z|^{2\nu+2} e^{n(1-\mu^2)(z^2+{z^*}^2)/4\mu^2}
    K_\nu\left (\frac{n(1+\mu^2)|z^2|}{2\mu^2} \right).
  \ee 
The integral diverges logarithmically when one of the eigenvalues is close
to $z$. While the divergent term dominates the partition function, the divergence
can be absorbed into the normalization.
Then the bosonic determinant cancels against the same factor from the Vandermonde determinant and the partition function reduces to \cite{AHOV,kim-bos}
\be
Z_{0/1+1^*}(z,z^*,\mu) \sim w(z,z^*;\mu) \log(\epsilon).
\ee
This gives rise to a  baryon density and a chiral condensate that depend smoothly
on the chemical potential and the phase transition of the the fermionic theory
at $\mu =m_\pi/2$ does not take place.

The logarithmic singularity is a generic feature of the bosonic partition
function which can also be understood starting from a chiral Lagrangian.
The Dirac operator in the phase quenched bosonic partition function has
to be regularized as \cite{janik-hermitization}
\be
D^{\rm reg} = \left(\begin{array}{cc}
 \hat D+ \mu\gamma_0  & \epsilon\\
  -\epsilon  & -\hat D +\mu\gamma_0\end{array}\right )
\label{d-reg}
\ee
with the chiral block structure of the Dirac operator $\hat D$ given by
\be
\hat D = \bmat m & id \\ id^\dagger  &m \emat.
\ee
The determinant of this two-flavor Dirac operator can be rewritten as
\be
\det ( (\hat D+\mu \gamma_0)^\dagger(\hat D+\mu \gamma_0) +\epsilon^2) =
\det ( \hat D \ident + \mu \gamma_0 \tau_3 + \epsilon \gamma_5 i \tau_2),
\ee
so that, physically, $\epsilon$ is the source term for the isospin condensate.
By permutation of rows and columns,  the regularized determinant
operator can be written as
\be
\det \tilde D^{\rm reg}
\ee
with
\be
 \tilde D^{\rm reg}=
  \left(\begin{array}{cc}
 \epsilon +im\tau_2 & id\tau_1+\mu i\tau_2\\
 id^\dagger\tau_1+\mu i\tau_2  & \epsilon+im\tau_2  \end{array}\right ),
\label{d-reg2}
\ee
which makes it possible to express
the bosonic partition function as a convergent Gaussian integral
\be
Z_{0/1+1^*}=\left \langle \int \prod_k d\phi_k d\phi_k^*
\exp \left  [ -\vect \phi_1^* \\ \phi_2^* \evect^T
  \left(\begin{array}{cc}
 \epsilon +im\tau_2 & id\tau_1+\mu i\tau_2\\
 id^\dagger\tau_1+\mu i\tau_2  & \epsilon+im\tau_2  \end{array}\right )
  \vect \phi_1 \\ \phi_2 \evect \right ] \right \rangle.
  \ee
  The pion condensate  is given 
  by the expectation value
  \be
  \frac 1{n}\langle \phi_1^*\cdot \phi_1+ \phi_2^*\cdot \phi_2 \rangle,
\label{pi-cond}
  \ee
which follows by differentiation with respect to the source term.
  A nonzero value of this condensate spontaneously breaks the symmetry 
the Gl(1)/U(1) symmetry 
  \be
  \vect \phi_1 \\ \phi_2 \evect \to
  \exp( s\tau_3) \vect \phi_1 \\ \phi_2 \evect, \qquad
  \vect \phi_1^* \\ \phi_2^* \evect^T \to
  \vect \phi_1^* \\ \phi_2^* \evect^T\exp( s\tau_3),
  \label{gl1}
   \ee
   of $\tilde D^{\rm reg}$ 
   with $ s $ real (for $\epsilon \to 0 $).
   Note that an imaginary part of $s$ would violate
   the complex conjugation property of the integration variables and the
   integral would no longer be convergent. 
   In the chiral Lagrangian, the $s$-degree of freedom becomes a ``Goldstone
   mode'' which for nonzero $\epsilon$ acquires a mass term
   \be
   \sim \epsilon \Tr e^{\tau_3 s} = 2\epsilon \cosh s.
\label{s-mass}
   \ee
   The integral over $s$ gives the
$ \log \epsilon$-divergence of the partition
   function found earlier in this section. This is  a general argument that applies both to the ensemble
   $D_1$ and the ensemble $D_2$ and applies as long as the above Gl(1)/U(1) symmetry
   is spontaneously broken.

   The source term for the chiral condensate is the quark mass, and it is thus given by
  \be
  \frac 1{n}\langle \phi_1^* i\tau_2 \phi_1+ \phi_2^* i\tau_2 \phi_2 \rangle.
  \ee
  The corresponding Goldstone manifold for the noncompact symmetry is thus given by
  \be
  e^{s\tau_3} \Sigma_c e^{s\tau_3}
  \ee
with
   $\Sigma_c = i\tau_2$.
  The $s$ degree of freedom drops out of the Goldstone manifold, and it is not
  possible to regularize the partition function by introducing a regulator mass in this source term. If the partition
  function has to make sense we necessarily need a nonzero pion condensate  for which the
  Gl(1)/U(1) symmetry is spontaneously broken, and the Goldstone degree of freedom $s$
  acquires the mass term \eref{s-mass}.

  Let $m_c$  be the critical mass such that for $m<m_c$, $m$ is inside the support of the
  spectrum of $\hat D$, while for $m>m_c$  it is outside of this region. Then it is clear
  that  the anti-Hermitian Dirac operator \eref{d-reg2} does not have
   a gap for $m<m_c$ (as a function of $\epsilon$), and the  symmetry \eref{gl1} is spontaneously
   broken. For $m>m_c$, although the spectrum of the matrix in \eref{d-reg2}  acquires
   a gap, the pion condensate \eref{pi-cond} remains nonzero. The reason is that
   the contribution of single eigenvalue of $\hat D +\mu \gamma_0$ close to the mass
   diverges as $\log \epsilon$ in the regularized partition function. This follows by
   writing  the phase quenched
   bosonic partition function
    in terms of the eigenvalues of the Dirac operator $\hat D +\mu \gamma_0$ as
   \be
   Z &=& \int \frac {\rho(\lambda_1,\cdots \lambda_n)}
   {\prod_{k=1}^n |m^2 -\lambda_k^2|^2}
   \prod_{k=1}^n d \lambda_k d \lambda_k^*\nn \\
   & \sim & \frac{n\log \epsilon}{4m^2} 
   \int \frac {\rho(\lambda_1,\cdots \lambda_{n-1},\pm m)}
        {\prod_{k=1}^{n-1} |m^2-\lambda_k^2|^2}
        \prod_{k=1}^{n-1} d \lambda_k  d\lambda_k^*\nn. \\
\label{z-eps}
 \ee
For the partition function $D_2$ the bosonic determinant cancels against
the Vandermonde determinant, and we find that the chiral condensate is
given by $m/\mu^2$. For the partition function $D_1$ it is not possible
to further simplify \eref{z-eps}, but we expect that the chiral condensate
remains continuous at $m = m_c$.
Indeed, for
 the random matrix ensemble $D_1$, the partition function is still
dominated by the logarithmic singularity due to a single eigenvalue close
to the quark mass, and because of eigenvalue repulsion, there are no other eigenvalues
close to $m$. In particular, the joint eigenvalue density
$\rho(\lambda_1,\cdots \lambda_{n-1},\pm m)$ vanishes linearly for any of the $\lambda_1, \cdots, \lambda_{n-1}$
close  to $\pm m$.
However,  we no longer have the exact cancellation of the
bosonic determinant against the Vandermonde determinant.

The chiral Lagrangian for the phase quenched partition function of $D_1$ was derived in \cite{split-fact}. The mean
field limit of the corresponding partition function given by (in units where $\Sigma =1$)
\be
Z(m,\mu) = e^{-4n\mu^2 -nm^2/\mu^2}
\label{zpq-mft}
   \ee
   results in the chiral condensate
   \be
   \Sigma(m,\mu) = -\frac 1{2n} \frac d{dm} \log Z(m,\mu) = \frac m{\mu^2},
   \ee
   and the baryon density 
   \be
   n_B(\mu) -\frac 1{2n} \frac d{d\mu} \log Z(m,\mu) = 4\mu-\frac{m^2}{\mu^3}.
   \ee
   In the Bose-condensed phase the mean field limit of the fermionic phase
   quenched partition function is given by
   \be
   Z_{1+1^*/0}(m,\mu) = e^{4n\mu^2 +nm^2/\mu^2}
     \ee
     resulting in the same chiral condensate and baryon density as obtained
     for the bosonic partition function.
     In the normal phase ($m>2\mu^2$) the mean-field limit of the phase quenched partition
     function is given by
     \be
     Z_{1+1^*/0}(m,\mu) = e^{4nm}.
   \ee
   This phase is not present in the bosonic partition function.
   
What we learn from this example is that in order to obtain the $\log \epsilon$
dependence, the noncompact flavor symmetry of the bosonic partition function
has to be broken spontaneously. If it would not be broken, the noncompact
degree of freedom could not be regularized and the regularization that works
for the fundamental theory, would fail for the effective theory.

\section{One Flavor Partition Function at Imaginary Chemical Potential}

The fermionic one-flavor partition function of the random matrix theory $D_1$
was analyzed in \cite{andy,lehner}
for imaginary chemical potential and in \cite{misha,HJV} for real chemical
potential. Some of the relevant results for the fermionic partition function will
be reviewed in the next subsection, while the bulk of this section is devoted to the
derivation of an analytical expression for the bosonic partition function, and a comparison
of observables for the two partition functions.

\subsection{The Fermionic Partition Function at Nonzero (Imaginary) Chemical Potential}

The fermionic one-flavor partition function can be evaluated by writing the
determinant as a Grassmann integral and performing a Hubbard-Stratonovitch
transformation after averaging over the randomness, or alternatively by
super-bosonization
\cite{efetov-original,martin,efetov,akemann-basile,super-mario}.
The exact result for finite $n$ in the sector of topological charge $\nu$ is given by
\cite{andy,HJV}
\be
Z^\nu(m,\mu)= \int_0^\infty ds s^{\nu+1}  I_\nu(2 m n s\Sigma) (s^2-\mu^2)^n e^{-n\Sigma^2(s^2-\mu^2+m^2)}.
\ee
This result is valid both for arbitrary complex chemical potential, and in particular for real or purely imaginary chemical potential. It has two phases, a chirally broken phase and a phase with restored chiral symmetry. In units where $\Sigma =1$, the critical curve
is given by \cite{andy,misha,HJV}
\be
Re(1+\mu^2+\log \mu^2) = 0.
\ee
In Fig. 1 we show this curve in the complex $\mu$-plane. The first order
lines end at $\mu =\pm i$ where the transition is of second order.
\begin{center}
\begin{figure}
 \centerline{ \includegraphics[width=7cm]{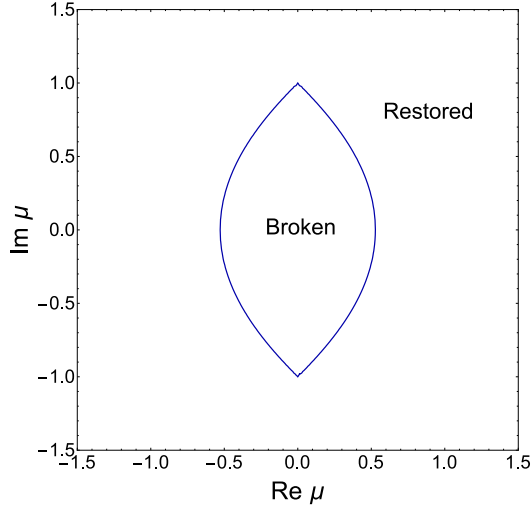}}
  \caption{Phase diagram of the random matrix partition function in the
    complex $\mu$ plane in units where the chiral condensate is
    equal to 1.}
\end{figure}
\end{center}

An alternative expression for the fermionic partition function can be obtained
by means of the superbosonization technique. The result can be expressed as
(see Appendix \ref{app-A})
\be 
   {Z^\nu(m,\mu) =\frac{(n+1)! e^{n\Sigma^2\mu^2}}{(n+1-\nu)!}
     \int_{-\pi}^\pi d\beta  \int_{-1}^1 xdx
e^{-i\beta( 2n+\nu)} x^\nu  I_\nu(2me^{i\beta}x )J_0(2\mu e^{i\beta}\sqrt{1-x^2} )
e^{ e^{2i\beta} x^2/n\Sigma^2}.\hspace*{2cm}}\hspace*{-2.4cm}\nn\\
\ee
The integrals over $x$ and $\beta$ can be performed analytically resulting
in a finite sum that can easily be evaluated numerically.

\subsection{The Bosonic Partition Function}

After averaging over the chiral random matrix ensemble, 
the one-flavor bosonic partition function for $\nu =0$
and imaginary chemical potential is given by \cite{sener1}
\be
{\hspace*{-0.5cm}Z(m,i\mu_i)= e^{\nb\Sigma^2 \mu_i^2} \int d\phi^*_1d\phi_1d\phi^*_2d\phi_2 
\exp\left  [ i \vect \phi_1^* \\ \phi_2^* \evect 
\mat i m & \mu_i \\ \mu_i & i m \emat
\vect \phi_1 \\ \phi_2 \evect 
-\frac {\phi_1^*\cdot \phi_1 \phi_2^*\cdot \phi_2}{\nb\Sigma^2}  \right ],
\hspace*{1cm}}\hspace*{-1cm}
\ee
where the normalization factor $\exp {\nb\Sigma^2 \mu^2}$ is chosen
to give a $\mu$-independent partition function in the chiral limit
below the critical point.
We distinguish $\nb$ appearing in the probability distribution and
the number of components $n$ of the vector $\phi_1$.  
Instead of using a Hubbard-Stratonovitch transformation to linearize
the quartic term, we use the bosonic part of the superbosonization
transformation to evaluate the integral.
The starting point is to insert the $\delta$ -function
\be
\delta (\Phi -S)
\ee
in the partition function
with $S$ a positive definite Hermitian matrix
and 
\be
\Phi = \mat \phi_1^*\cdot \phi_1 &  \phi_1^*\cdot \phi_2 \\
              \phi_2^*\cdot \phi_1 &  \phi_2^*\cdot \phi_2 \emat.
\ee
The partition function can then be rewritten as
\be
Z(m,i\mu_i) = e^{\nb\Sigma^2\mu_i^2} \int dS d\Phi \delta (S- \Phi)
e^{-m\tr \Phi  +i\mu_i \tr \sigma_1 \Phi - S_{11} S_{22}/\nb\Sigma^2},
\ee
where the integral is over Hermitian matrices $S$.
The $\delta$-function can be expressed as \cite{Fyodorov}
\be
 \delta (S- \Phi) = \int dF e^{i \tr F(S-\Phi -i\epsilon)} 
\ee
resulting in the partition function
\be
Z(m,i\mu_i) = e^{\nb\Sigma^2 \mu_i^2} \int dSd d\Phi dF e^{iFS} 
e^{-m\tr \Phi  +i\mu_i \tr \sigma_1 \Phi-i\tr F\Phi   - S_{11} S_{22}/\nb\Sigma^2}.
\ee
The integral over $\Phi$ evaluates to
\be
Z(m,i\mu_i) = e^{n\mu_i^2} \int dS dF e^{i \tr FS}  
\frac 1{\det^n[F-im -\sigma_1 \mu_i]}
e^{   - S_{11} S_{22}/n\Sigma^2}.
\label{eq7}
\ee 
The integral over $F$ is an Ingham-Siegel integral 
\cite{ingham,siegel,Fyodorov,split-fact} which is known analytically,
\be
\int dF {\det}^{-n}(F-i\epsilon) e^{i \tr SF} = 
\theta(S) {\det}^{n-2} S e^{-\epsilon \tr S},
\ee
where $\theta(S)$ indicates that $S$ is positive definite.
We thus find
\be
Z(m,\mu_i) = e^{\nb\Sigma^2 \mu_i^2} \int_{S>0} dS {\det}^{n-2}S  
e^{-m(S_{11}+S_{22})+i\mu_i(s_{12}+s_{21})   - S_{11} S_{22}/\nb\Sigma^2}.
\ee

For $\nu \ne 0$, we choose $\phi_1$ to be of length $n+\nu$ and $\phi_2$ of
length $n$.  When comparing 
different topological sectors \cite{moshe}, we will put $\bar n = n +\nu/2$
and keep $\bar n$ fixed so that the number
of eigenvalues of the Dirac matrix is the same for different $\nu$.
In Eq. (\ref{eq7}) this results in an extra factor $1/(F_{11}-iz)^\nu$,
\be
Z^\nu(m,i\mu_i) = e^{\nb\Sigma^2 \mu_i^2} \int dS e^{i \tr FS}  
\frac 1{\det^n[F-im -\sigma_1 \mu_i](F_{11}-im)^\nu}
e^{   - S_{11} S_{22}/\nb\Sigma^2},
\ee 
and after shifting the diagonal matrix elements of $F$ by $im$, we need to evaluate the integral
\be
\int dF {\det}^{-n}(F-i\epsilon) (F_{11}-i\epsilon)^{-\nu} e^{i \tr SF}.
\label{boson-integral}
\ee
To calculate this integral we rewrite the determinant to obtain
\be
\int dF {(F_{22}-i\epsilon-F_{21}F_{12}/(F_{11}-i\epsilon}))^{-n} (F_{11}-i\epsilon)^{-\nu-n} e^{i \tr SF}.
\ee
The integral over $F_{22}$ can be performed by a contour integration resulting
in
\be
\frac{2\pi i^n}{(n-1)!}S_{22}^{n-1} \theta(S_{22})\int dF
     s(F_{11}-i\epsilon)^{-\nu-n} 
e^{i \tr  S_{22}\frac{F_{21}F_{12}}{F_{11}-i\epsilon} +iS_{11}F_{11} +iS_{12}F_{21}+iS_{21}F_{12} }.\nn\\
\ee
The integral over $F_{12}$ and $F_{21}=F_{12}^*$ is a Gaussian integral 
which can be easily evaluated. We find
\be
\frac{2\pi i^n}{(n-1)!}S_{22}^{n-1} \frac{\pi i}{S_{22}}\theta(S_{22})\int dF  (F_{11}-i\epsilon)^{-\nu-n+1} 
e^{-i \tr  S_{12}S_{21}F_{11}/S_{22} +iS_{11}F_{11}  }.
\ee
Also the integral over $F_{11}$ can be performed by a contour integration so
that we finally obtain for the integral \eref{boson-integral}
\be
&&-\frac{2\pi(-i)^{\nu+1}}{(n-1)!}S_{22}^{n-2} \pi  \frac{2 \pi }{(n+\nu-2)!} 
(S_{11}-S_{12}S_{21}/S_{22})^{n+\nu-2} 
\theta(S_{22})\theta(S_{11}-S_{12}S_{21}/S_{22})\nn \\
&=&-\frac{4(\pi (-i))^{nu+1}}{(n-1)!(n+\nu-2)!}{\det}^{n-2}S [\det S/S_{22}]^\nu  
\theta(S),
\ee
where $\theta(S) $ denotes that $S$ is positive definite.

The integration over positive definite matrices $S$ can be performed
by using the parameterization
\be
S = e^v \mat e^u \cosh s & i e^{i\phi} \sinh s \\
-ie^{-i\phi} \sinh s & e^{-u} \cosh s \emat.
\ee
The integration measure is given by
\be
dS = 4 e^{4v} \cosh s \sinh s.
\ee
This results in the partition function
\be
Z^\nu(m,i\mu_i) &=& \frac {e^{\nb\Sigma^2 \mu_i^2} }{(n-1)!(n+\nu-2)!} \int dv du ds d\phi  \cosh s \sinh s e^{2nv} 
\left[ \frac {e^{u+v}}{\cosh s} \right ]^\nu
\nn \\
&&\times
e^{-2me^v\cosh s\cosh u 
-2i\mu_i e^v \sinh s \sin \phi -e^{2v} \cosh^2 s/\nb\Sigma^2}.
\ee
The integrals over $u$ and $\phi$ can be expressed in terms of Bessel functions
\be
Z^\nu(m,i\mu_i) &=& \frac{e^{\nb\Sigma^2 \mu_i^2}}{(n-1)!(n+\nu-2)!}  \int dv ds\frac{ \cosh s \sinh s }{\cosh^\nu s }e^{(2n+\nu)v} 
K_\nu(2me^v \cosh s) J_0(2\mu_i e^v \sinh s)
\nn\\ &&\times
e^{ -e^{2v} \cosh^2 s/\nb\Sigma^2}.
\label{bos-res-1}
\ee

After shifting the $v$-integration by $\log\cosh s$ 
and choosing $x =\exp v$ as a new integration variable we obtain
 \be
{Z^\nu(m,i\mu_i) =\frac{ e^{\nb\Sigma^2\mu_i^2}}{(n-1)!(n+\nu-2)!}  \int dx ds\frac{ \cosh s \sinh s }{\cosh^{2n+2\nu} s }x^{2n+\nu-1} 
K_\nu(2mx) J_0(2\mu_i x \tanh s)
e^{ -x^2 /\nb\Sigma^2}.\hspace*{2cm}}\hspace*{-2.4cm}\nn \\
\ee
The integral over $y$ can be evaluated as a Bessel function resulting
in the expression
\be\hspace*{-0.5cm}
{Z^\nu(m,i\mu_i) = \frac 12 e^{\nb\Sigma^2 \mu_i^2} \frac{\mu^{1-n-\nu}\nb^{(n+1)/2}}{(n-1)!}
\int_0^\infty dx  x^{n} J_{n+\nu-1}(2\mu_i x\sqrt{ \nb})
K_\nu(2mx\sqrt{ \nb})e^{-x^2/\Sigma^2},\hspace*{2cm}}\hspace*{-2cm}
\label{zbos-bes}
  \ee
where we have also rescaled the integration variable by $\sqrt{ \nb}$.
This form can easily be evaluated numerically also for large values of $n$.
However, because of the oscillatory nature of the integrand, it is not
amenable to mean field estimates.

Next we derive an expression for the partition function in terms of
a positive definite integrand. This result
can be obtained
if we insert the following representation for the $K_\nu$ function
\be
K_\nu(x)= \frac 12 \frac {x^\nu}{2^\nu} \int_0^\infty \frac{ds}{s^{\nu+1}} e^{-s-x^2/4s}
\ee
resulting in
\be
Z^\nu(m,i\mu_i) &=& \frac 14  e^{\nb\Sigma^2 \mu_i^2}   \frac{\mu_i^{1-n-\nu}\nb^{(n+1)/2}}{(n-1)!}       
\int_0^\infty 
ds \frac{(x m \sqrt {\nb} )^\nu }{s^{\nu+1}} 
\int_0^\infty dx  x^{n} J_{n+\nu-1}(2\mu_i x\sqrt {\nb})
\nn\\ &&\times
e^{-s-m^2x^2\nb/s-x^2/\Sigma^2}.
\ee 
The integral over $x$ is known analytically \cite{ryzhik}
\be
\int_0^\infty dx x^{n+\nu} J_{n+\nu-1}(\beta x) e^{-\alpha x^2} 
= \frac {\beta^{n+\nu-1}}{(2\alpha)^{n+\nu}}e^{-\beta^2/4\alpha}.
\ee
After changing the integration variable be $s \to s \bar n m^2$ we find
\be
Z^\nu(m,i\mu_i) &=& \frac {{\bar n}^ne^{\nb\Sigma^2 \mu_i^2} } {8(n-1)!}               
m^{-\nu}\int_0^\infty \frac{ds}{s^{\nu+1}} e^{-s\bar n m^2}
\frac{1}{(1/s+1/\Sigma^2)^{n+\nu}} e^{-\nb \mu^2_i/(1/s+1/\Sigma^2)}\nn\\
&=& \frac {{\bar n}e^{n\Sigma^2 \mu_i^2} } {8}               
m^{-\nu}\int_0^\infty s^\nu \frac{ds}{s} e^{-\bar n m^2/s}
\frac{1}{(s+1/\Sigma^2)^{n+\nu}} e^{-\nb \mu^2_i/(s+1/\Sigma^2)}
,
\label{final-res}
\ee
where we also changed $s\to 1/s$ in the last line.
\subsection{Limiting Cases}

In this subsection,  we derive three limiting cases of  \eref{final-res},
the microscopic limit, the $\mu_i\to 0$ limit, the chiral limit and
 the large $n$-limit of the bosonic partition function.

In  the microscopic limit for the mass, $m \nb = {\rm fixed}$ for $\nb \to \infty$ and $n\to \infty$ with $n \to \bar n$
at fixed imaginary chemical potential the partition function
simplifies to
\be
Z^\nu(m,i\mu_i) &=& \frac {e^{n\Sigma^2 \mu_i^2} }{8(n-1)!}        (\nb\Sigma^2)^{n+\nu}
m^\nu\int_0^\infty \frac{ds}{s^{\nu+1}} e^{-s}
e^{-\nb^2m^2\Sigma^2/s -\mu^2_i\nb\Sigma^2 (1 -\nb\Sigma^2m^2/s)}\nn\\
&=&\frac {e^{n\Sigma^2 \mu_i^2} }{8(n-1)!}         (\nb\Sigma^2)^{n+\nu}
m^\nu
2 (\nb m\Sigma \sqrt{1-\mu^2_i\Sigma^2})^{-\nu} e^{ -\mu^2_i\nb\Sigma^2 }
K_\nu(2\nb m\Sigma\sqrt{1-\mu^2_i\Sigma^2})\nn\\
&=&\frac {e^{n\Sigma^2 \mu_i^2} }{8(n-1)!}   
\frac{\nb^{n} \Sigma^{2n+\nu}}
{(1-\mu^2_i\Sigma^2)^{\nu/2}} e^{ -\mu^2n\Sigma^2 }
K_\nu(2\nb  m\Sigma\sqrt{1-\mu^2_i \Sigma^2}),
\ee
which is consistent with the result obtained in \cite{sener1}.

 For $\mu_i=0$ the partition function \eref{bos-res-1}  can be written as
\be
Z^\nu(m,i\mu_i=0) = \int_0^\infty dx ds  x^{2n+\nu-1} \cosh s \sinh s
K_0(2mx \cosh s)
e^{ -x^2 \cosh^2 s/\nb\Sigma^2}.
\ee
After rescaling $x$ by $\cosh s$ , the $s$ integral gives an overall constant so that the
partition function simplifies to
\be
Z^\nu(m,i\mu_i=0) = \int_0^\infty dx   x^{2n+\nu-1} 
K_\nu(2mx )
e^{ -x^2 /\nb\Sigma^2}.
\ee
This is indeed the Cauchy transform of a Laguerre polynomial \cite{ake-fyo}, which is
the correct finite $n$ result for the chiral random matrix partition function.

For $m \to 0$ we have that
\be
K_\nu(2mx\sqrt{ \nb}) &\sim &\frac 12 (mx \sqrt nb)^{-\nu} \qquad {\rm for} \quad \nu \ne 0,\nn \\ 
K_0(2mx\sqrt{ \nb})& \sim & -\log m  \qquad {\rm for} \quad \nu = 0. 
\ee
For $\nu=0$, the chiral limit can be worked out analytically
\be
Z^{\nu=0}(m\to 0,i\mu_i) = -\frac 12 e^{n\Sigma^2 \mu_i^2} \frac{\mu^{1-n}\nb^{(n+1)/2}}{(n-1)!}\log m
\int_0^\infty dx  x^{n} J_{n-1}(2\mu_i x\sqrt{ \nb})
e^{-x^2/\Sigma^2}.
\label{zbos-bes2}
  \ee
  This integral is known analytically \cite{ryzhik} resulting in
\be
Z^{\nu=0}(m\to 0,i\mu_i) = -
\frac{\nb^{n} 2^{2n-2}}
     {(n-1)! \Sigma^{2n}}\log m .
\label{zbos-bes3}
  \ee
  In the chiral limit, the partition function is dominated by the logarithmic singularity which does not depend on the imaginary chemical
  potential.  Contrary to the fermionic partition function, it
  is always in a phase with zero ``baryon density''. 
  

For large $n$ the partition function can be evaluated by a saddle point approximation. 
The saddle point equation for the expression in the second line of \eref{final-res} reads
\be
-\frac {m^2}{s^2} + \frac 1{1+s} - \frac{\mu^2}{(1+s)^2} =0.
\ee
To leading order in $m$ the solution is given by
\be
\bar s = \left \{ \begin{array}{c}
\displaystyle
\frac m{\sqrt{1-\mu^2_i}}, \quad \mu_i < 1,\\
\displaystyle
\mu^2_i-1, \quad \mu_i >1,
\end{array}
\right . 
\ee
resulting in the free energy ($\bar F = (\log Z )/n$)
\be
\bar F = \left \{ \begin{array}{c}
\displaystyle
 2m{\sqrt{1-\mu^2_i}}, \quad \mu_i < 1,\\
\displaystyle
 1-\mu^2_i+\log \mu^2_i + \frac {m^2}{\mu^2_i-1}, \quad \mu_i >1.
 \end{array} \right . 
 \ee
 The chiral condensate is given by
   \be
-\frac 1{2n}\frac{d\log Z}{dm} = \left \{ \begin{array}{c}
\displaystyle
    {\sqrt{1-\mu^2_i}}, \quad \mu_i < 1,\\
    \displaystyle
 \frac {m}{\mu^2_i-1}, \quad \mu_i>1,
\end{array} \right .
 \ee
      and the baryon number density by
 \be
- \frac 1{2n}\frac{d\log Z}{d\mu_i} =
 \left \{ \begin{array}{c}
   \displaystyle
   0 , \quad \mu_i < 1,\\
   \displaystyle
 -\frac 1{\mu_i} + \mu_i, \quad \mu_i >1 .
 \end{array} \right . 
 \ee
 The baryon number susceptibility  at imaginary chemical potential is  defined
 by
\be
\chi_B = -\frac 1{2n}\frac {d^2 \log Z}{d\mu^2_i}=
\left \{ \begin{array}{c}
\displaystyle
 0 , \quad \mu_i < 1,\\
\displaystyle
 \frac 1{\mu_i^2} + 1, \quad \mu_i >1 .
 \end{array} \right . 
\label{suscep}
  \ee
 \begin{figure}[t!]
\includegraphics[width=8cm]{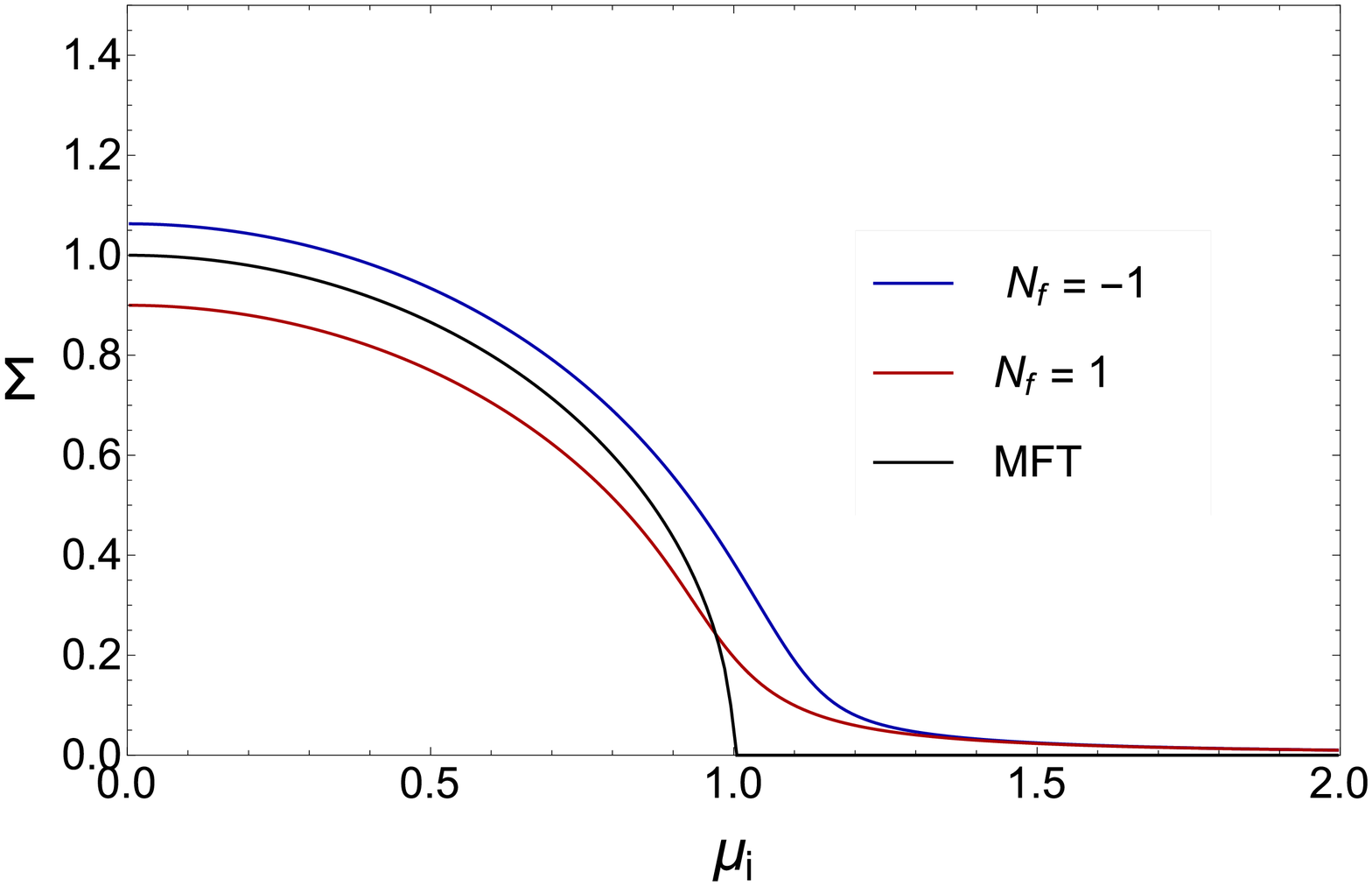}
\includegraphics[width=8cm]{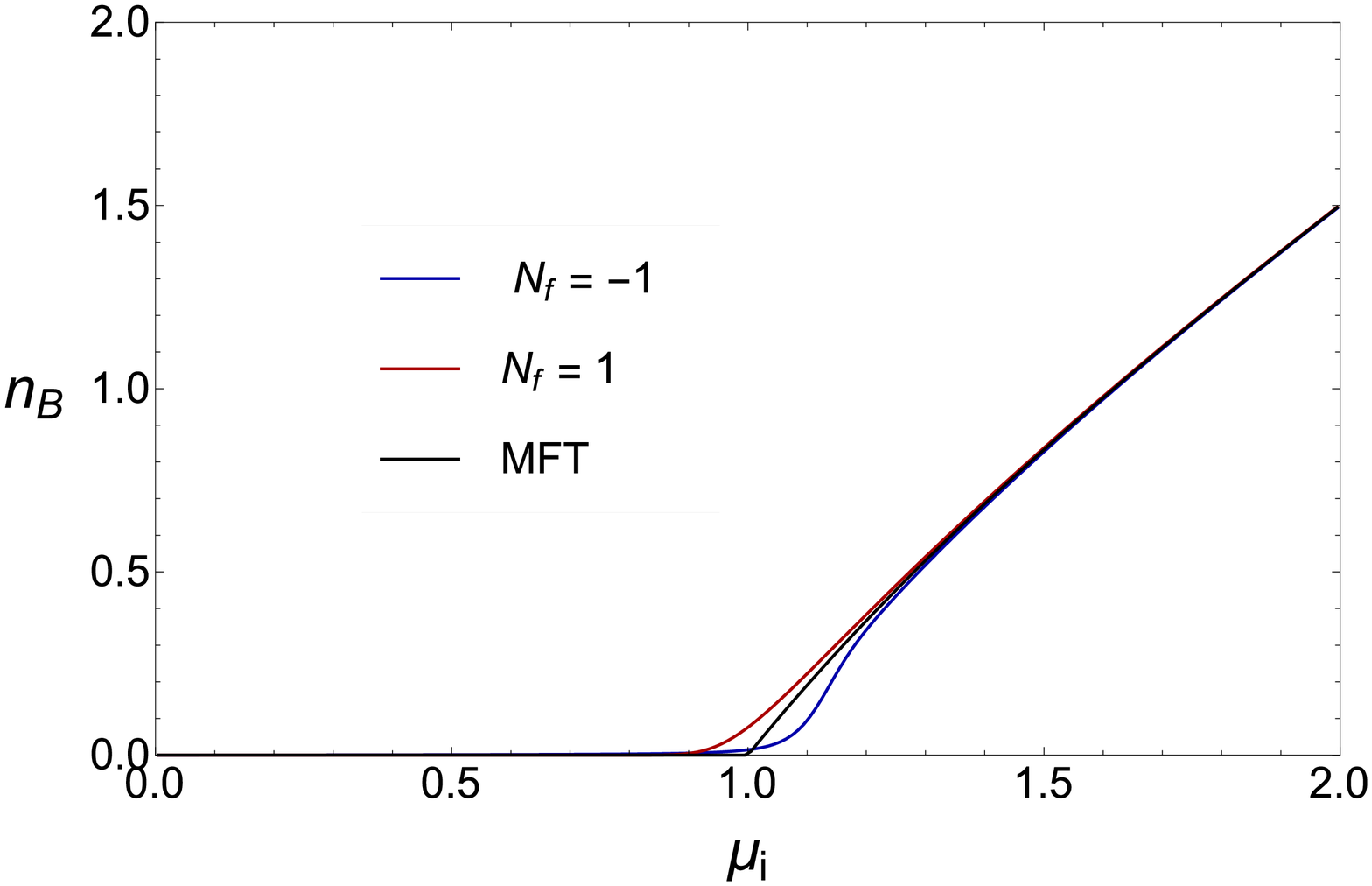}
\caption{ The chiral condensate (left) and the baryon density (right)
   as a function of the
  imaginary chemical potential. We show the result for the one-flavor bosonic partition
  function (blue), the one flavor fermionic partition function (red) and the mean
field result (black).}
\label{fig:fig1}
 \end{figure}

 In Fig. 2 we show the chiral condensate (left) and the baryon number (right)
 as a function of the imaginary chemical potential. The results are for
 $n=100$, $m =3/100$ in case of the chiral condensate and $n=100$, $m=3/10000$ in
 case of the baryon number
 all in  units with $\Sigma =1$ in the partition function. Both the results
 for the fermionic partition function (blue) and the bosonic partition function (red)
 are close to the mean field result (black) which has been obtained for
 $n\to \infty$ in the chiral limit. 

\begin{center}
 \begin{figure}[b!]
\centerline{\includegraphics[width=8cm]{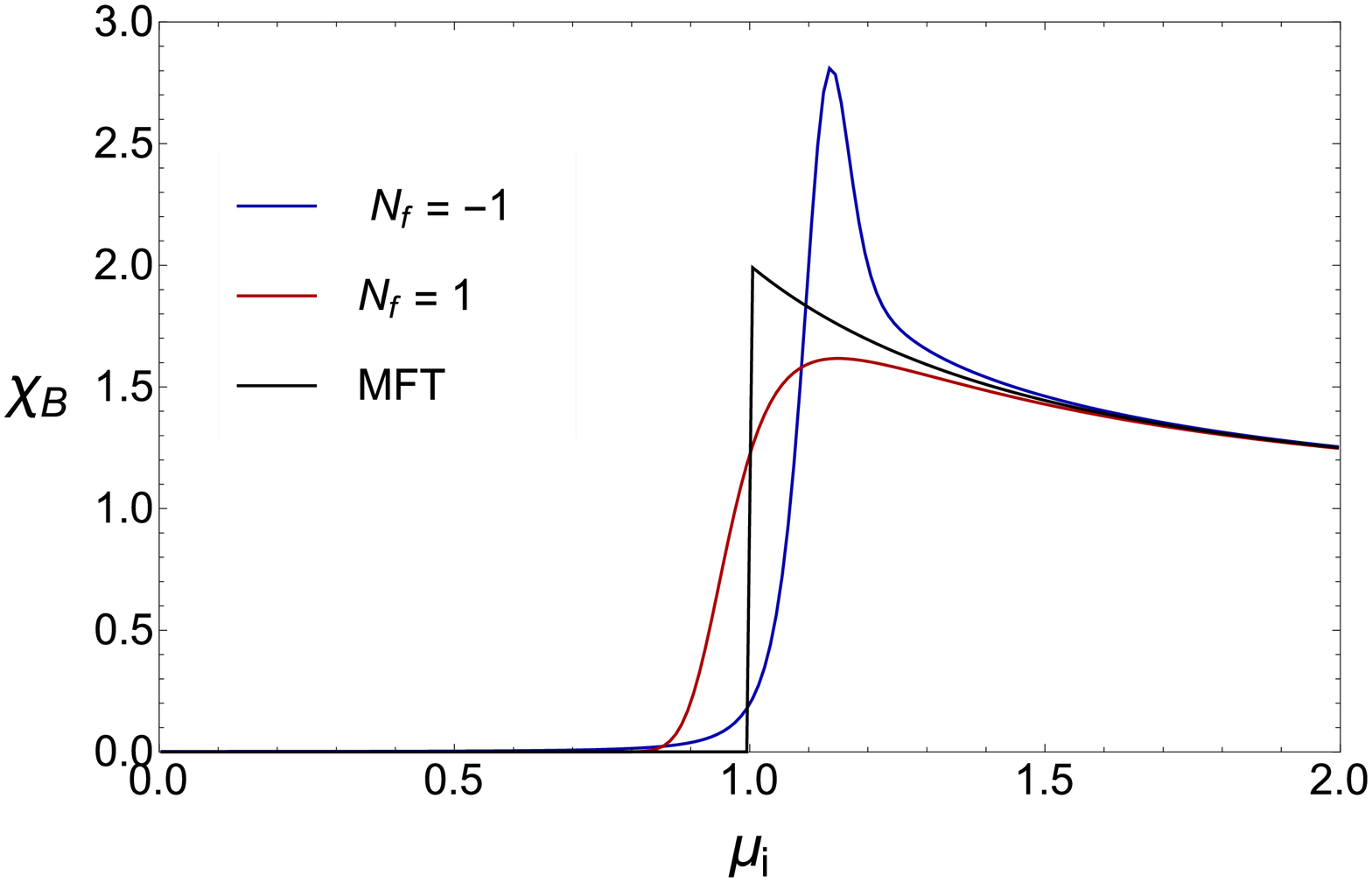}}
\caption{ The baryon number susceptibility as a function of the imaginary
  chemical potential, $\mu_i$ for $n=100$ and $m=3/10000$. Results are shown for the
  fermionic partition function (red), the bosonic partition function (blue), and the
   mean field limit of these partition functions.}
\label{fig:sus1}
 \end{figure}
\end{center}

The baryon number susceptibility defined in Eq. \eref{suscep} is shown
in Fig. \ref{fig:sus1} as a function of the imaginary chemical potential
for $n=100$ and $m=3/10000$. Again
the bosonic and fermionic susceptibility are close to the mean field result, but the
deviation near the critical point is much larger than in  case of the
baryon number density (see Fig. \ref{fig:fig1}). The convergence of the
susceptibility to the thermodynamical limit is non-uniform in $m$.

 \section{Bosonic Partition Function for Real Chemical Potential}

 In this section we consider the massless bosonic chiral random matrix partition function for real
 chemical potential.  In this case, the partition function can be expressed in terms
 of the joint probability distribution of the Ginibre ensemble, which allows
 us to obtain exact analytical results. We start with a heuristic derivation
 of the mean field results for the chemical potential dependence of the partition
 function, and in the second subsection we reduce this partition function to
 a two-dimensional integral. Everywhere in this section we work in units where $\Sigma =1 $ and
 in the sector of zero topological charge.
  
 \subsection{Heuristic Derivation of the Mean Field Result}
\label{heur}

In units where $\Sigma = 1$ and $\nu =0$, the massless bosonic partition function can be expressed as
\be
Z_{0/1}(\mu) = e^{-n\mu^2}\left \langle \det \frac 1{D(\mu)}\right \rangle
\ee
with $D(\mu)$ given by
\be
D(\mu) = \mat     0 & id + \mu \\ -id^\dagger +\mu & 0\emat ,
\ee
and the normalization factor $\exp(-n\mu^2)$ has been included to give the correct $\mu$ dependence
for small $\mu$.
If $\mu$ is inside the domain of eigenvalues, the partition function 
has to be regularized. This can be done in the same way as for the
phase quenched bosonic partition function,
\be
Z_{0/1}(\mu) = \left \langle \frac {{\det}^* D(\mu)}
{\det (D(\mu) D(\mu)^\dagger + \epsilon^2)} \right \rangle,
\ee
where the limit $\epsilon \to 0$ has to be taken at the end of the calculation.
Contrary to the partition function with a pair of conjugate bosonic quarks at nonzero chemical potential,
this partition function, because of the extra fermionic determinant, is finite for $\epsilon \to 0$.
At the mean field level we expect that this partition function is given by the ratio of two fermionic
partition functions,
\be
Z_{0/1}^{\rm MFT}(\mu) =\frac{Z_{N_f=1}(\mu)}{Z_{N_f=1+1^*}(\mu)},
\ee
where $Z_{N_f=1+1^*}(\mu)$ is the phase quenched partition function, or equivalently, the
product of the same one flavor partition function and
the bosonic  phase quenched partition function (see Eq.  \eref{zpq-mft}). 
The baryon density is thus given by
\be
n_B &=& - \frac 1{2n}\frac d{d\mu} \log Z_{0/1}(\mu) \nn\\
&& \frac 1{2n}\frac d{d\mu} \log Z_{1+1^*}(\mu) -\frac d{d\mu} \log Z_{1}(\mu) .
\ee
The $ \mu $ dependence of both partition functions is well known
\cite{split-fact,misha} and is given by
\be
\frac 1{2n} \frac d{d\mu} \log Z_{1+1^*}(\mu)
&=&\theta(1-\mu) 4\mu +\theta(\mu-1) (2\mu + \frac 2\mu)
\nn \\
\frac 1{2n} \frac d{d\mu} \log Z_{1}(\mu)
&=&\theta(\mu-\mu_c) (\mu + \frac 1\mu).
\ee
   \begin{figure}[t!]
\includegraphics[width=8cm]{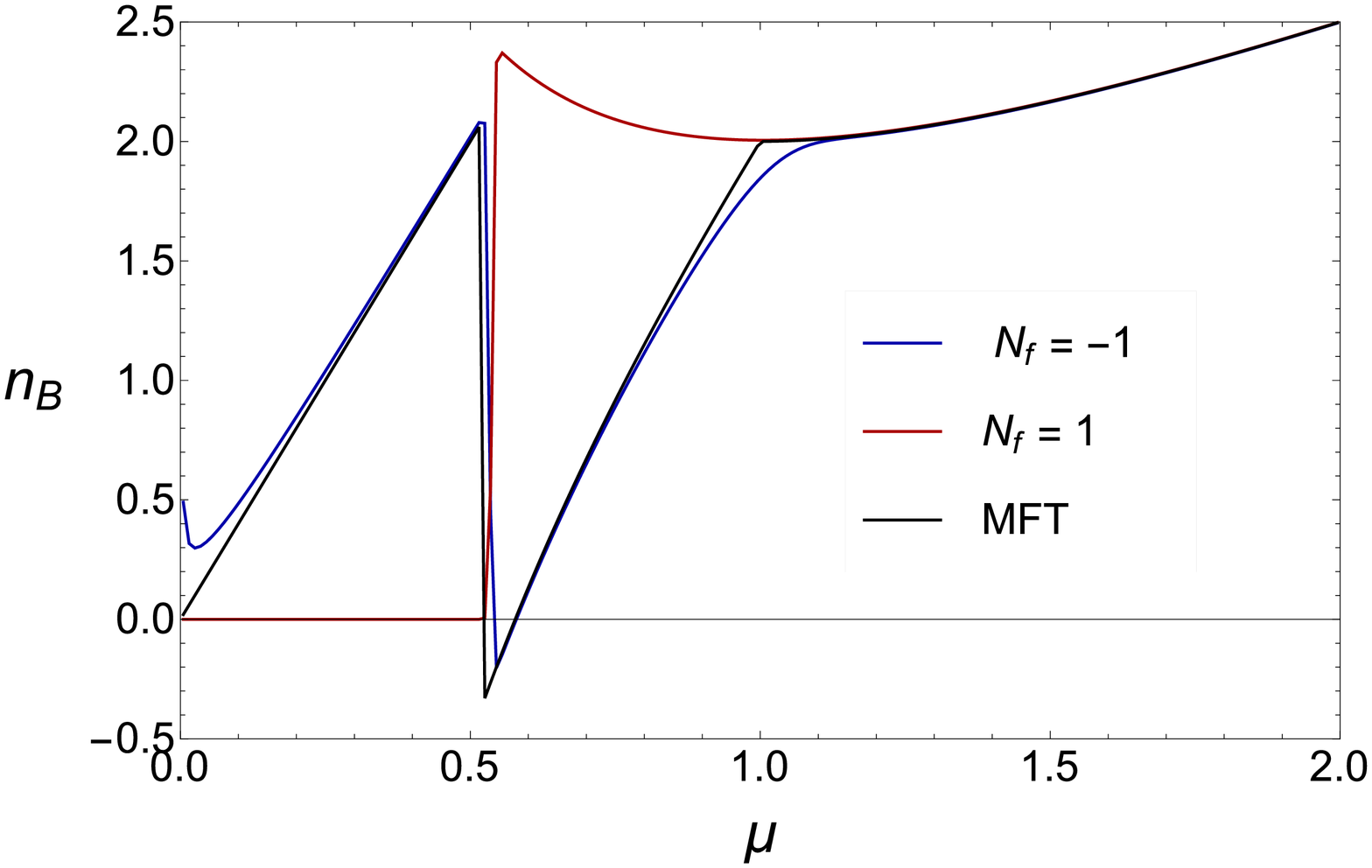}
\includegraphics[width=8cm]{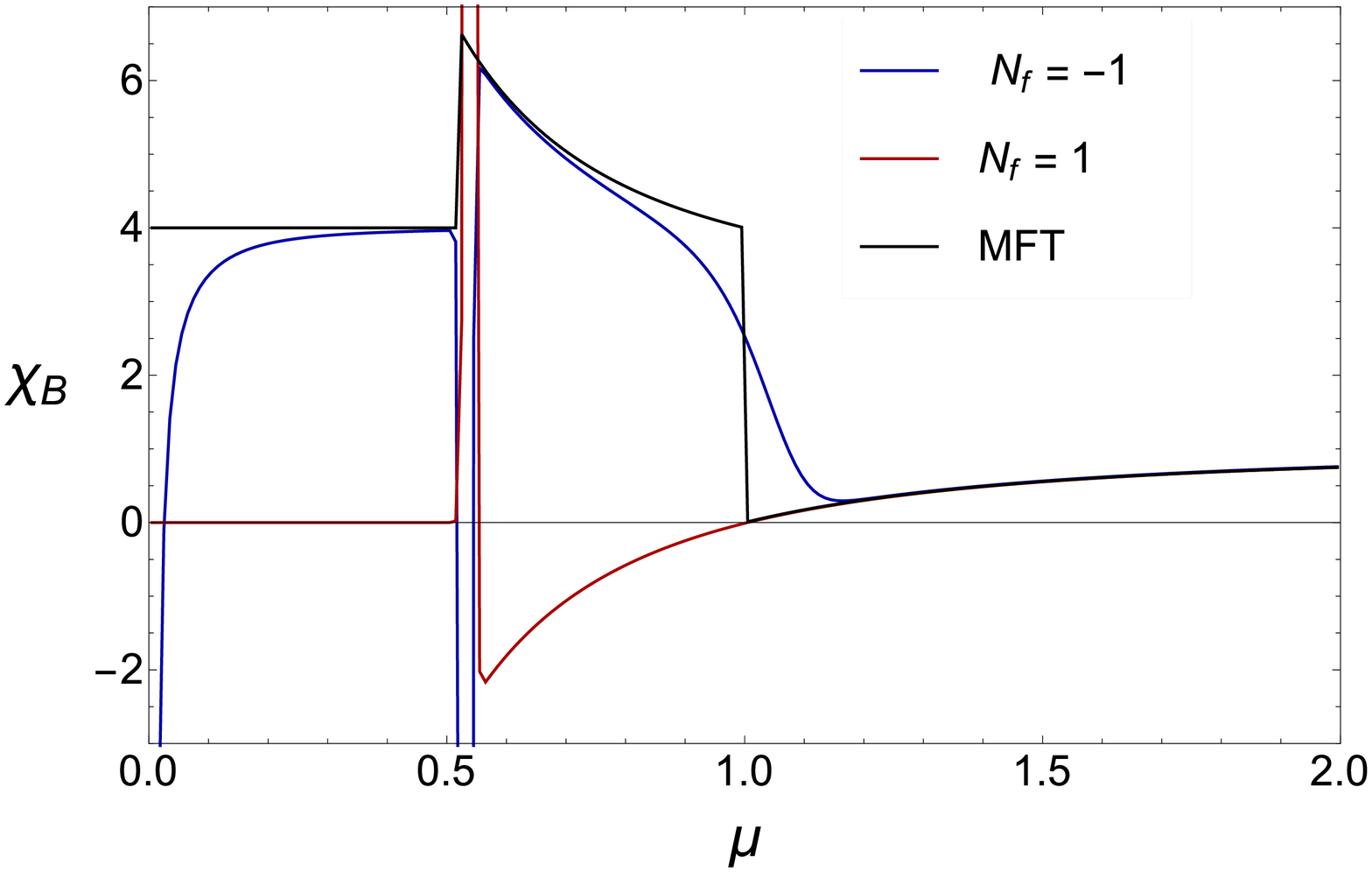}
\caption{ The baryon number density (left) and baryon number susceptibility (right) as a function of the 
  chemical potential, $\mu$ for $n=100$ and $m=0$. Results are give for the
  fermionic partition function (red), the bosonic partition function (blue), and the
   mean field limit of the bosonic partition function.}
\label{fig:susm1}
 \end{figure}
where $\mu_c =0.527$. In Fig. \ref{fig:susm1}, the black curve represents the mean field result for the baryon density. In the same figure we have plotted
the analytical for finite $n$ (blue curve), which will be derived in the
next subsection, and the finite $n$ result for the baryon density of the fermionic partition function (red curve).
When $\mu$ is outside the domain of eigenvalues, the fermionic and bosonic results become equal in the thermodynamic
limit.

\subsection{The Finite $n$ Massless Bosonic Partition Function at Nonzero Chemical Potential}
\label{bosmu}

In this section we evaluate the massless bosonic random matrix partition function as a function
of the real baryon chemical potential. This partition function can be written as
(the equality only holds for even $n$) \cite{ipsen-split}
\be
\left \langle \frac 1{\det (d + \mu)\det(-d^\dagger+\mu)}\right \rangle
= \left \langle \frac 1{\det (d + \mu)\det(d^\dagger -\mu)}\right \rangle,
\ee
where the matrix elements of the complex 
$n\times n$ matrix $d$  are distributed according to
\be
p(d) = e^{- \nb {\rm Tr} d^\dagger d} .
\ee 
The quenched matrix ensemble with this distribution, known as the
Ginibre ensemble, has the
joint eigenvalue  density 
\be
P(z_k) = \Delta(\lambda_k) \Delta(\lambda^*_k) \prod_k e^{-\nb \lambda_k^* \lambda_k},
\ee
where $\Delta(\lambda_k)$ is the Vandermonde determinant.
The corresponding monic orthogonal polynomials and their normalization are equal to
\be
p_n(z) = z^n,\quad{\rm with}\quad h_n=\int dz dz^* p_n^*(z) p_m(z)= \delta_{nm}\frac{n!}{\nb^{n+1}}.
\ee
The partition function of the Ginibre ensemble, defined as the integral over the
probability distribution, can be obtained by expressing the Vandermonde determinants
in terms of these orthogonal polynomials. Performing the integrals by means of  orthogonality
relations we obtain
\be
Z_n^{\rm G} = n! \prod_{k=0}^{n-1} h_k.
\label{Ginibre}
\ee

In terms of the eigenvalues of $d$, the bosonic partition function can be written as
\be
Z_{0/1}(\mu) = \frac {e^{-n\mu^2}}{Z_n^{\rm G}}
\int \prod_k d\lambda_k d\lambda_k^* \frac{e^{-\nb\lambda_k^* \lambda_k}}{(\lambda_k+\mu)(\lambda_k^*+\mu)}
| \Delta(\lambda_k)|^2.
\ee
To evaluate this partition function we need
 identity
\be
u^{n-1} \frac {\Delta_k(\lambda_i)}{\prod_{k=1}^n(\lambda_k -u)}
= \sum_{k=1}^n(-1)^{k+n}
\frac{\lambda_k^{n-1} }{(\lambda_k-u)}
{\Delta'_k(\lambda_i)},
  \ee
  where
\be
\Delta'_p({z_k})= \prod_{k<l, kl \ne p} (z_k-z_l).
\ee
This identity can be proved by
including the factors $1/(\lambda_k -u)$ in the determinant and
expanding it with respect to the last row.
 Applying this identity to the  bosonic determinant results in
\be
(-1)^{n-1}\mu^{2n-2} \frac 1{\prod_{k=1}^n(\lambda_k -u)(\lambda_k^*+u)} = \sum_{k,l=1}^n
(-1)^{k+l}
\frac{\lambda_k^{n-1} {\lambda_l^*}^{n-1}}{(\lambda_k-\mu)(\lambda_l^*+\mu)}
\frac{\Delta'_k(\lambda_i)\Delta'_l(\lambda_j^*)}
{\Delta_k(\lambda_i)\Delta_l(\lambda_j^*)}.
\ee 
We can distinguish two types of terms, those with $k=l$, and those with
$k\ne l$. All terms of each type give the same contribution to the
bosonic partition function. We thus find
\be
Z_{0/1}(\mu)& =&  e^{-n\mu^2}
\frac {n(-1)^{n-1}}{Z_n^{\rm G} \mu^{2n-2}}\int\prod_k  d\lambda_k d\lambda_k^*e^{-\nb\lambda_k\lambda_k^*}
\frac {(\lambda_1 \lambda_1^*)^{n-1}}{(\lambda_1-\mu)(\lambda_1^*+\mu)}
\Delta'_1(\lambda_i)\Delta'_1(\lambda_j^*) \nn\\
&&
-  e^{-n\mu^2}\frac {n(n-1)(-1)^{n-1}}{Z_n^{\rm G} u^{2n-2}}\int d\lambda_k d\lambda_k^*
e^{-\nb\lambda_k\lambda_k^*}
\frac {(\lambda_1 \lambda_2^*)^{n-1}}{(\lambda_1-u)(\lambda_2^*+u)}
\Delta'_1(\lambda_i)\Delta'_2(\lambda_j^*),\nn\\
\ee
where the partition function is normalized with respect to the Ginibre partition function
\eref{Ginibre}.
This expression can be rewritten as
\be
Z_{0/1}(\mu)& =&  e^{-n\mu^2}\frac{ n(-1)^{n-1}}{Z_n^{\rm G} \mu^{2n-2}}\int d\lambda_1 d\lambda_1^*e^{-\nb\lambda_1\lambda_1^*}
\frac {(\lambda_1 \lambda_1^*)^{n-1}}{(\lambda_1-u)(\lambda_1^*+u)}
Z_{n-1}^{\rm G}
 \nn\\
&&
+  e^{-n\mu^2}\frac {n(n-1)(-1)^{n}}{Z_n^{\rm G} \mu^{2n-2}}\int d\lambda_1 d\lambda_1^*d\lambda_2 d\lambda_2^*
e^{-\nb(\lambda_1\lambda_1^*-\lambda_2\lambda_2^*)}
\frac {(\lambda_1 \lambda_2^*)^{n-1}}{(\lambda_1-\mu)(\lambda_2^*+\mu)}
\nn\\ &&\times
\langle \pi_{n-2}(\lambda_2)\pi_{n-2}(\lambda_1^*)\rangle Z_{n-2}^{\rm G}.
 \ee
 where the average of two characteristic polynomials is defined by
\be
\langle \pi_{n-2}(u)\pi_{n-2}(v^*)\rangle =
\frac {1}{Z_{n-2}^{G}} \int \prod_{k=1}^{n-2} \frac{d\lambda_k d\lambda_l^*}\pi
  \prod_{k=1}^{n-2} (u-\lambda_k)(v^*-\lambda_k^*)e^{-\nb\lambda_k\lambda_k^*}
    |\Delta(\lambda_1,\cdots, \lambda_{n-2})|^2 .\nn\\
    \ee
    This average can be expressed in terms of the two-point kernel of the
    Ginibre ensemble \cite{Akemann:2002vy}
    \be
    \langle \pi_{n-2}(u)\pi_{n-2}(v^*)\rangle =
    h_{n-2}\sum_{k=0}^{n-2}\frac{(uv^*)^k}{h_k}.
    \ee
    This results in the partition function
\be
Z_{0/1}(\mu) & =&  \frac {(-1)^{n-1}e^{-n\mu^2}}{h_n \mu^{2n-2}}\int d\lambda_1 d\lambda_1^*e^{-\nb\lambda_1\lambda_1^*}
 \frac {(\lambda_1 \lambda_1^*)^{n-1}}{(\lambda_1-\mu)(\lambda_1^*+\mu)}
 \\
&&
-  \frac {(-1)^{n-1}e^{-n\mu^2}}{h_{n-1} \mu^{2n-2}}\int d\lambda_1 d\lambda_1^*d\lambda_2 d\lambda_2^*
e^{-\nb\lambda_1\lambda_1^*-\nb\lambda_2\lambda_2^*}
\frac {(\lambda_1 \lambda_2^*)^{n-1}}{(\lambda_1-\mu)(\lambda_2^*+\mu)}
\sum_{k=0}^{n-2}\frac{(\lambda_2 \lambda_1^*)^k}{h_k}.\nn
\label{zu1}
\ee
 This derivation is also valid for complex values of $\mu$. The first integral
 in Eq. \eref{zu1} is logarithmically divergent for purely imaginary $\mu$ and has to be regularized which can be done by including a mass term. 
 The resulting logarithmically divergent part of the partition function is
 $\mu$ independent, which agrees
with the result for the chiral limit of the bosonic partition function at imaginary chemical
potential which diverges as $\log m$ for $\nu = 0$ (see Eq. \eref{zbos-bes3}).

The integrals can be calculated using polar coordinates and converting
the angular integral to a contour integral,
\be
Z_{0/1}(\mu)
&=&
\frac {(-1)^{n-1}e^{-n\mu^2}}{h_{n-1} \mu^{2n-2}\pi}\int d\lambda d\phi e^{-\nb\lambda^2}
 \frac {\lambda^{2n-2}}{(\lambda e^{i\phi} -\mu)(\lambda e^{-i\phi}+\mu)}
 \nn\\
&&
-  \frac {(-1)^{n-1}e^{-n\mu^2}}{h_{n-1} \mu^{2n-2}\pi^2}
\sum_{k=0}^{n-2} \frac{(-1)^{k+n}}{h_k} \left[\int d\lambda d\phi
e^{-\nb\lambda^2}
\frac {\lambda^{n+k}e^{i\phi (n-1-k)}}{(\lambda e^{i\phi}-\mu)}
\right ]^2
\nn \\
 &=&
\frac {2 (-1)^{n-1}e^{-n\mu^2}}{h_{n-1} \mu^{2n-2}}\left[- \int_0^{|\mu|} d\lambda  e^{-\nb\lambda^2}
 \frac {\lambda^{2n-1}}{\mu^2+\lambda^2}
+\int_{|\mu|}^\infty d\lambda  e^{-\nb\lambda^2}
 \frac {\lambda^{2n-1}}{\mu^2+\lambda^2}\right ]
 \nn\\
&&
-  \frac {4(-1)^{n-1}e^{-n\mu^2}}{h_{n-1} \mu^{2n-2}}
\sum_{k=0}^{n-2} \frac{(-1)^{k+n}}{h_k}  \left[\int_{|\mu|}^\infty d\lambda 
e^{-\nb\lambda^2}
\lambda^{2k+1}\mu^{n-2-k}
\right ]^2
\nn \\
 \label{z-bos-1}
 \ee
 Note that this partition function is not an analytic function of $\mu$ which was
 also the case for the bosonic partition function of  model \eref{D2} \cite{kim-bos}. 
Because of large cancellations this form of the partition function is not amenable
to a mean field analysis. In Appendix \ref{app:B} we derive a form   where
these cancellations have been taken care of analytically. It is given by (for $\mu>0$)
\be
Z_{0/1}(u)&=&
\frac{e^{-n\mu^2}}{h_{n-1}}\left [
 \int_{0}^\infty dx 
\frac{e^{-\nb\mu^2(2x+1)}}{x+1} 
 - \int_{0}^1 dx
 \frac {e^{-\nb\mu^2(2x+1)}}{x+1}
 \left ( \frac{\Gamma(n-1,-\nb\mu^2x)}{\Gamma(n-1)}\right )
\right . \nn \\ &&\left .
 +\int_{0}^1 dx 
 \frac{e^{-\nb\mu^2(x+y)}}{x+1}(-x)^{n} \left(1-\frac{\Gamma(n-1,\nb \mu^2}{\Gamma(n-1)}
\right)
 \right ].
\label{zmu-final}
\ee
We have checked that this result agrees with a direct evaluation of
the partition function for $n=2$ and $n=3$. See Appendix \ref{app:C} for the brute force
expressions for $n=2$ and $n=3$.
   \begin{figure}[t!]
\centerline{\includegraphics[width=8cm]{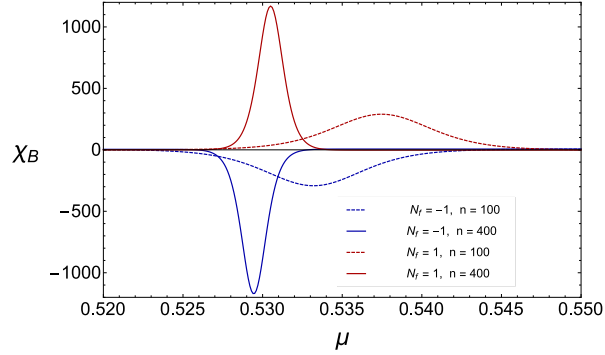}}
\caption{ The baryon number susceptibility near the critical point for $\nb=n=100$
  (dashed) and $\nb=n=400$ (solid) both for the bosonic 9 (blue) and the fermionic
  partition function. Up to a minus sign, which could have been absorbed by the
  definition of the bosonic baryon number susceptibility, the result are similar.
}
\label{fig:53}
 \end{figure}

\subsection{Large $n$ Limit of the Bosonic Partition Function}
  
In the large $n$ limit, where we take $\nb =n$, the first term of Eq. \eref{zmu-final} is given by
\be
\frac 1{h_{n-1}} \frac 1{4n u^2} e^{-4n\mu^2}
\sim \frac {e^{-n(4\mu^2-1)}}{4\mu^2\sqrt{2\pi n}},
\ee
and the second term by
\be
\left \{ \begin{array}{c}
    \displaystyle
    \frac 1{h_{n-1}\Gamma(n-1)}
    \frac {e^{-3n \mu^2}(-n\mu^2)^{n-3}}{2 n\mu^2(1+1/\mu^2)(1/\mu^2-1)}
       \sim \frac{(-1)^n e^{-n(3\mu^2-\log \mu^2 -2)}}
          {4\pi n^2 \mu^2(1+\mu^2)(1-\mu^2)}
,\qquad {\rm for} \quad \mu < 1  \\
\displaystyle 
\frac 1{h_{n-1}} \frac {e^{-2n\mu^2}}{2n u^2(1+1/\mu^2)} 
\sim \frac{e^{-n(1+2\mu^2)}}{\sqrt{2\pi  n}  2n(1+\mu^2)}
\qquad {\rm for}\quad \mu > 1 \quad
\end{array} \right . .
\ee
The last term factorizes into the product of two integrals. For large $n$
it can be approximated by
\be
\left \{ \begin{array}{c}
\displaystyle
  \frac {\Gamma(n-1)}{h_{n-1}} \frac{(-1)^ne^{-n\mu^2}}{(n\mu^2)^n (1+1/\mu^2)}
\sim \frac{(-1)^n \mu^2 e^{-n\log \mu^2}-n\mu^2}{1+\mu^2}
\qquad {\rm for} \quad  u>1\\
\displaystyle
\frac 1{h_{n-1}\Gamma(n-1)} \frac {(-1)^n e^{-3n\mu^2}(-n\mu^2)^{n-3}}
      {2 (1-1/\mu^2)^2}\sim \frac{(-1)^ne^{-n(3\mu^2-\log \mu^2 -2)}}{4\pi n \mu^2(1-\mu^2)^2}
\qquad {\rm for}\quad  \mu<1  \end{array} \right .
  \ee
  This result agrees with the heuristic estimate of section \ref{heur}.

   In Fig. \ref{fig:susm1} we show the baryon number density and the baryon number susceptibility as a function of the chemical potential for $n=100$ and $m=0$. Results are given
   for fermionic partition function (red), the bosonic partition function (blue) and
   the mean field limit of the bosonic partition function. The susceptibility diverges
   at $\mu=\mu_c =0.523$ as $\sim n$ in the thermodynamical limit, see Fig. \ref{fig:53}.
   This reflects that the slope
   \be
   \left .  \frac {d n_B}{d\mu} \right |_{\mu=\mu_c} \sim n.
\ee
Note that we could have defined the baryon number susceptibility with the opposite sign.

   \section{Conclusions}

   We have studied bosonic random matrix partition functions (averages of inverse
   determinants)
   and compared them to fermionic random matrix partition functions (averages of
   a determinants) for the same value of the external parameters. In particular, we consider
   the dependence of the chiral condensate, the baryon density and the baryon number susceptibility on the (imaginary) chemical
   potential and the quark mass. For imaginary chemical potential, $\mu_i$, and nonzero quark mass, these observables approach the same limit
   for $n\to \infty$, where the $\mu_i$-dependence is given by the mean field result
   of the effective partition function.
   In the chiral limit, the bosonic partition function diverges as $\log m$ whereas the fermionic partition function remains finite.

   We have seen two cases where the bosonic partition is always in the broken phase while
   the fermionic partition function undergoes a phase transition to the restored phase.
   The first case is the phase quenched partition function, where the pion condensate
   of the bosonic partition function is nonvanishing for all $\mu$ while it is becomes zero
   for $\mu < m_\pi/2$ in case of the fermionic partition function. The second case is
   the chiral limit of the one flavor partition function as a function of imaginary chemical potential. In this case the
   fermionic partition function has as phase transition to the restored phase at
   $\mu_i =1$ while the bosonic partition diverges as $\log m$ and is in the same
   phase for all values of $\mu_i$. As a side remark we note that this gives us
   two more examples where the replica
   trick is doomed to fail \cite{critique}.

   The spontaneous breaking of noncompact symmetries has also been studied for
   hyperbolic spin models in one and two dimensions. The conclusion of this work
   is that a noncompact symmetry is always
   broken spontaneously, even in one and two dimensions, 
   if the partition function diverges for vanishing symmetry
   breaking term. Our work supports
   this conclusion for a different class of models.

\section*{Acknowledgments}

This work was supported by  U.S. DOE Grant No. DE-FAG-88FR40388. Gernot Akemann, Kim Splittorff and Mario Kieburg are thanked for useful discussions. In particular, we thank Kim Splittorff
for the suggestion to study the model of section IV.


\appendix

\section{Derivation of the Fermionic Partition Function Using Superbosonization}
\label{app-A}

Superbosonization was developed as an alternative to the Hubbard-Stratonovitch
transformation \cite{efetov-original,martin,efetov,akemann-basile,super-mario}
in order to be
able to deal with non-Gaussian probability distributions.
Below we only use the fermion-fermion part of the superbosonization transformation. The fermionic partition function is given by

\be
Z^\nu(z,\mu)= e^{n\Sigma^2\mu^2}\int d\chi^*_1d\chi_1d\chi^*_2d\chi_2 
\exp\left  [  \vect \chi_1^* \\ \chi_2^* \evect 
\mat  z & \mu \\ \mu &  z \emat
\vect \chi_1 \\ \chi_2 \evect 
+\frac 1{n\Sigma^2} \chi_1^*\cdot \chi_1 \chi_2^*\cdot \chi_2 \right ],
\hspace*{0.5cm}
\ee
where the vector $\chi_1$ is of length $n+\nu$ and the length of the vector $\chi_2$ is
of length $n$. 
To linearize the four-fermion term, we use the fermion-fermion part of the
superbosonization transformation by inserting
the $\delta$-function
\be
\delta(G-Y) = \int dF e^{i\tr F(G -Y)}
\ee
with
\be
G =\mat \chi_1^*\chi_1 &\chi_1^*\chi_2\\
\chi_2^*\chi_1 & \chi_2^*\chi_2 \emat,
\ee
and $Y^\dagger = Y$.  After integration of the $\chi$-variables,
this results in the partition function,
\be
{Z^\nu(z,\mu)= e^{n\Sigma^2\mu^2}\int dY dF F_{11}^\nu {\det}^n F
\exp\left  [ z(Y_{11}+Y_{22})+\mu(Y_{12}+Y_{21})   
+\frac {Y_{11}Y_{22}}{n\Sigma^2}  -i\tr FY
\right ].\hspace*{2cm}}\hspace*{-2cm}\nn\\
\ee 
The integral over $F$ can be evaluated by means of an Itzykson-Zuber
integral as
\be
[i\del_{Y_{11}}]^\nu  \int dF {\det}^n F e^{ -i\tr FY} 
\ee
with 
\be
\int dF {\det}^n F e^{ -i\tr FY} &=&
\int \prod df_k \Delta^2(\{f_k\}) \prod f_k^n
\frac{\det e^{if_ky_l}}{\Delta(\{f_k\})\Delta(\{y_l\})}\nn\\ 
&=& \frac {\det \delta^{(n+k-1)} (y_l)}{\Delta(\{y_l\})}
\ee
where $\delta^{(p)}(y)$ is the $p$-th derivative of a $\delta$-function.
Acting on a regular test function $F(Y)$, it has the property
\cite{akemann-basile}
\be
\int DY \frac {\det \delta^{(n+k-1)} (y_l)}{\Delta(\{y_l\})}  F(Y) 
& =& \int dY \sum_\pi (-1)^{\sigma_\pi}
\left . \frac{\prod_{k=1}^p\del_{y_{\pi(k)}}^{n+k-1}F(Y)}{\Delta(\{y_l\})}\right |_{y_k = 0}
\nn \\
& =& \int dU  \oint \prod_k{dy_k}{2\pi i}   \Delta^2(\{y_l\})
 \sum_\pi (-1)^{\sigma_\pi}\frac {F(Y)}{\Delta(\{y_l\}) \prod_{k=1}^p y_{\pi(k)}^{n+k}}\nn\\  
& =& \int dU  \oint \prod_k\frac{dy_k}{2\pi i}  \Delta^2(\{y_l\}) 
 \sum_\pi (-1)^{\sigma_\pi}\frac {F(Y)\prod_{k=1}^p y_{\pi(k)}^{p-k}}{\Delta(\{y_l\}) \prod_{k=1}^p y_k^{n+p}}
 \nn\\  
& =& \int_{Y\in U(p)} \frac{dY}{(2\pi)^4} {\det}^{-n-p} Y F(Y),  
\ee
where in the second last equation we have used that the last product is
a Vandermonde determinant. Note the measure $dY$ is the product over independent
differentials.
We thus arrive at the partition function
\be
Z^\nu(z,\mu)&=& e^{n\Sigma^2\mu^2}\int_{Y\in U(2)} \frac{dY}{2\pi^4} [[i\del_{Y_{11}}]^\nu {\det}^{-n-2} Y]
\exp\left  [ z(Y_{11}+Y_{22})+\mu(Y_{12}+Y_{21})   
+\frac {Y_{11}Y_{22} }{n\Sigma^2} 
\right ] \nn \\
&=& \frac{(n+\nu+1)!e^{n\Sigma^2\mu^2}}{(n+1)!}\int_{Y\in U(2)} \frac{dY}{(2\pi)^4} \frac{[-iY_{22}]^\nu}{ {\det}^{n+2+\nu} Y} \\ &&\times
\exp\left  [ z(Y_{11}+Y_{22})+\mu(Y_{12}+Y_{21})   
+\frac 1{n\Sigma^2} Y_{11}Y_{22} 
\right ].\nn
\ee  
We parameterize $Y$ as
\be
Y = e^{i\beta} \mat e^{i\alpha} \cos \theta  &  e^{i\phi} \sin \theta \\
          -e^{-i\phi}\sin \theta &   e^{-i\alpha} \cos \theta \emat,
          \ee
          where $\beta \in [0,2\pi]$, $\phi \in [0,2\pi]$, $\alpha \in [0,\pi]$ and
          $\theta \in [0,\pi]$.
The invariant measure is given by
\be
\frac{dY}{(2\pi)^3\det^2 Y} =  \frac{\cos\theta \sin\theta d\beta d\alpha d\theta d\phi}{2\pi^3}.
\ee 
This results in the partition function
\be
Z^\nu(z,\mu) &= &\frac{(n+\nu+1)!e^{n\Sigma^2\mu^2}}{2\pi(n+1)!}\int  \frac{d\beta d\alpha d\theta d\phi}{(2\pi)^2}  \cos\theta \sin\theta
e^{-2i\beta(n+\nu)}  (-i)^\nu e^{i\nu(\beta-\alpha)} \cos^\nu\theta
\nn\\&&\times
e^{2ze^{i\beta} \cos\theta \cos\alpha +2i\mu e^{i\beta}\sin \theta \sin\phi +
\frac 1{n\Sigma^2} e^{2i\beta} \cos^2\theta}.\nn\\
\ee
The integral over $\alpha$ and $ \phi$ gives a modified Bessel function
so that we finally obtain
\be
Z^\nu(z,\mu) &=& \frac{(n+\nu+1)!e^{n\Sigma^2\mu^2}}{2\pi (n+1)!}\int   \frac{d\beta  d\theta}{4\pi} \cos\theta \sin\theta
e^{-2i\beta n} e^{-i\beta\nu}\cos^\nu \theta I_\nu(2ze^{i\beta} \cos\theta)
\nn\\ &&\times
J_0(2\mu e^{i\beta} \sin \theta )
e^{\frac 1{n\Sigma^2} e^{2i\beta} \cos^2\theta}.\nn\\
\ee
The normalization will be fixed by the result for $\mu=0$.
\be
Z^\nu(z,\mu=0) &=& \frac{(n+\nu+1)!}{(n+1)!}\int   d\beta  d\theta \cos\theta \sin\theta
e^{-2i\beta n} e^{-i\beta\nu}\cos^\nu \theta I_\nu(2ze^{i\beta} \cos\theta)
e^{\frac 1{n\Sigma^2} e^{2i\beta} \cos^2\theta}\nn\\
&=& \frac{(n+\nu+1)!}{(n+1)!} \int   d\beta  dx x
e^{-2i\beta n} e^{-i\beta\nu}x^\nu I_\nu(2ze^{i\beta} x)
e^{\frac 1{n\Sigma^2} e^{2i\beta} x^2}\nn \\
&=&\frac{(n+\nu+1)!}{(n+1)!} \int_0^1 dx x
\sum_{k+l=n} (zx)^{2k+\nu} x^\nu (\frac 1{n\Sigma^2}  x^2)^l
\frac1{k!(k+\nu)! l!}\nn \\
&=&\frac 12 \frac{(n+\nu)!}{(n+1)!} 
\sum_{k=0}^n (z)^{2k+\nu} ({n\Sigma^2} )^{k-n}
\frac1{k!(k+\nu)! (n-k)!}.
\ee
The sum is exactly the expression for a Laguerre polynomial so that
\be
Z(z,\mu=0) &=& \frac {z^\nu}{(n+1)!\Sigma^{2n}}L_n^\nu(-z^2n\Sigma^2).
\ee
In the microscopic limit this reduces to
\be
Z(z,\mu=0) &=& \frac 1{(n+1)!\Sigma^{2n+\nu}} I_\nu(2z n\Sigma).
\ee
To get the correct $\nu$ dependence we have to include an additional
factor of $\Sigma^\nu$ in the partition function which was already observed
in \cite{lehner}.

\section{Massless one Flavor Bosonic Partition Function}
\label{app:B}
             The goal of this appendix to derive a form of the massless bosonic
             one flavor partition function where the cancellation of the
             leading order terms has been take care of analytically. The starting
             point is in the expression in \eref{z-bos-1}
\be
             Z_n^{N_f=-1}(\mu)    &=&
           \frac {(-1)^{n-1}}{h_{n-1} }\left[- \int_0^{1} dx  e^{-\nb \mu^2 x}
 \frac {x^{n-1}}{x+1}
+\int_{1}^\infty dx e^{-\nb\mu^2x}
 \frac {x^{n-1}}{x+1}\right ]
 \nn\\
&&
-  \frac {(-1)^{n-1}}{h_{n-1} }\sum_{k=0}^{n-2} \frac{(-1)^{k+n}}{h_k} 
\left[\mu^{k+1}\int_{1}^\infty dx
e^{-\nb\mu^2x}
x^{k}
\right ]^2.
\nn \\\label{part1}
\ee         
The sum on the second line of this equation can be written as
\be
&&\sum_{k=0}^{n-2} \frac{(-1)^{k}}{h_k} 
\left[\mu^{k+1}\int_{1}^\infty dx
e^{-\nb\mu^2x}
x^{k}
\right ]^2\nn\\
&&= 
\sum_{k=0}^{n-2} \frac{(-1)^{k}}{h_k} \mu^{2(+1)}
\int_{1}^\infty dy e^{-\nb\mu^2y} y^{k}(\int_{0}^\infty dx e^{-\nb\mu^2x} x^{k}
-\int_{0}^1 dx e^{-\nb\mu^2x} x^{k})
\nn\\
&&= 
\sum_{k=0}^{n-2} \frac{(-1)^{k}}{h_k} \mu^{2(k+1)}
\int_{1}^\infty dy e^{-\nb\mu^2y} y^{k}\left (\frac{k!}{(\nb\mu^2)^{k+1}}
-\int_{0}^1 dx e^{-\nb\mu^2x} x^{k}\right )\nn\\
&&= 
\sum_{k=0}^{n-2} 
\int_{1}^\infty dy e^{-\nb\mu^2y} (-y)^{k}
-\frac{\mu^{2(k+1)}}{h_k} \int_{0}^1 dx e^{-\nb\mu^2x} x^{k}
\int_{1}^\infty dy e^{-\nb\mu^2y} (-y)^{k}
\nn\\
&&= 
\int_{1}^\infty dy e^{-\nb\mu^2y} \frac{1-(-y)^{n-1}}{1+y}
-\sum_{k=0}^{n-2}\frac{ \mu^{2(k+1)}}{h_k} \int_{0}^1 dx e^{-\nb\mu^2x} x^{k}
\int_{1}^\infty dy e^{-\nb\mu^2y} (-y)^{k}
\nn\\ 
&&= 
(-1)^{n}\int_{1}^\infty dy  \frac{y^{n-1}e^{-\nb\mu^2y}}{1+y}
+\int_{1}^\infty dy \frac{e^{-\nb\mu^2y} }{1+y}
\nn\\&&
-\sum_{k=0}^{n-2}\frac{ \mu^{2(k+1)}}{h_k} \int_{0}^1 dx e^{-\nb\mu^2x} x^{k}
\int_{1}^\infty dy e^{-\nb\mu^2x} (-y)^{k}
\nn\\ 
&&= 
(-1)^{n}\int_{1}^\infty dy  \frac{y^{n-1}e^{-\nb\mu^2 y}}{1+y}
+\int_{1}^\infty dy  \frac{e^{-\nb\mu^2y}}{1+y}
\nn \\&&
- \nb\mu^2\int_{0}^1 dx \int_{1}^\infty dy
e^{-\nb\mu^2(x+y+xy)} \frac{\Gamma(n-1,-\nb\mu^2xy)}{\Gamma(n-1)}
\nn\\ 
\ee
Inserting this result in  the partition function  \eref{part1} we  find
\be
Z(\mu) &=& \frac{1}{h_{n-1}}\left [-(-1)^{n-1}\int_0^{1} dx  e^{-\nb \mu^2 x}
 \frac {x^{n-1}}{x+1}
 +\int_{1}^\infty dx e^{-\nb\mu^2x} \frac{1}{1+x}
\right .  \nn \\ && \left .
- \nb\mu^2\int_{0}^1 dx \int_{1}^\infty dy
e^{-\nb\mu^2(x+y+xy)} \frac{\Gamma(n-1,-\nb\mu^2xy)}{\Gamma(n-1)}\right ]
 \nn\\  
&=&
{\frac{1}{h_{n-1}}\left [(-1)^{n}\int_0^{1} dx  
 \frac {x^{n-1}e^{-\nb \mu^2 x}}{x+1}
+\int_{1}^\infty dx  \frac{e^{-\nb\mu^2x}}{1+x}
- \nb\mu^2\int_{0}^1 dx \int_{1}^\infty dy  e^{-\nb\mu^2(x+y+xy)}
\right . \hspace*{2cm}}\hspace*{-2cm}\nn \\ && \left .
 - \nb\mu^2\int_{0}^1 dx \int_{1}^\infty dy
 e^{-\nb\mu^2(x+y+xy)}\left ( \frac{\Gamma(n-1,-\nb\mu^2xy)}{\Gamma(n-1)}-1\right )
 \right  ]
\nn\\
&=&
\frac{1}{h_{n-1}}\left [-(-1)^{n-1}\int_0^{1} dx  e^{-\nb \mu^2 x}
 \frac {x^{n-1}}{x+1}
+ \nb\mu^2\int_{1}^\infty dx \int_{1}^\infty dy
e^{-\nb\mu^2(x+y+xy)}
\right . \nn \\ && {\left .
 - \nb\mu^2\int_{0}^1 dx \int_{1}^\infty dy
 e^{-\nb\mu^2(x+y+xy)}\left ( \frac{\Gamma(n-1,-\nb\mu^2xy)}{\Gamma(n-1)}-1\right )
 \right ].\hspace*{2cm}}\hspace*{-4cm}
\ee 
Next we partial integrate the last term with respect to $y$. This results in
\be
Z(\mu)&=&
\frac{1}{h_{n-1}}\left [-(-1)^{n-1}\int_0^{1} dx  e^{-\nb \mu^2 x}
 \frac {x^{n-1}}{x+1}
+ \nb \mu^2\int_{1}^\infty dx \int_{1}^\infty dy
e^{-\nb \mu^2(x+y+xy)} \right .
\nn \\ && 
 - \int_{0}^1 dx
 \frac {e^{-\nb \mu^2(2x+1)}}{x+1}
 \left ( \frac{\Gamma(n-1,-\nb \mu^2x)}{\Gamma(n-1)}-1\right )
\\&&\left .
 + \frac{\nb\mu^2}{\Gamma(n-1)}\int_{0}^1 dx \int_{1}^\infty dy
 \frac{e^{-\nb\mu^2(x+y)}}{x+1}(-x)^{n-1} (y\nb \mu^2)^{n-2}
 \right ].\nn
\ee
When the upper limit of the $y$-integral in the last term is extended to
$[0,\infty]$ it is equal to $\Gamma(n-1)$ and cancels the first term. What remains
is the $y$-integral over $[0,1]$. We thus find
\be
Z(\mu)&=&
\frac{1}{h_{n-1}}\left [
 \nb\mu^2\int_{1}^\infty dx \int_{1}^\infty dy
e^{-\nb\mu^2(x+y+xy)} 
 - \int_{0}^1 dx
 \frac {e^{-\nb \mu^2(2x+1)}}{x+1}
 \left ( \frac{\Gamma(n-1,-\nb\mu^2x)}{\Gamma(n-1)}-1\right )
\right . \nn \\ &&\left .
 -\frac{\nb\mu^2}{\Gamma(n-1)} \int_{0}^1 dx \int_{0}^1 dy
 \frac{e^{-\nb\mu^2(x+y)}}{x+1}(-x)^{n-1} (y\nb \mu^2)^{n-2}
 \right ].
\ee
The integrals over $y$ can be performed analytically resulting in
\be
Z(u)&=&
\frac{e^{-n\mu^2}}{h_{n-1}}\left [
 \int_{0}^\infty dx 
\frac{e^{-\nb\mu^2(2x+1)}}{x+1} 
 - \int_{0}^1 dx
 \frac {e^{-\nb\mu^2(2x+1)}}{x+1}
 \left ( \frac{\Gamma(n-1,-\nb\mu^2x)}{\Gamma(n-1)}\right )
\right . \nn \\ &&\left .
 -\int_{0}^1 dx 
 \frac{e^{-\nb\mu^2x}}{x+1}(-x)^{n-1} \left(1-\frac{\Gamma(n-1,\nb\mu^2}{\Gamma(n-1)}
\right)
 \right ].
\label{zmu-final-app}
\ee

\section{Bosonic Partition function for $n=2$ and $n=3$}
\label{app:C}

In this appendix we evaluate the bosonic partition function without relying on
the tricks used in section \ref{bosmu}. Starting from the definition we obtain
given by
\be
Z_2(\mu)Z_2^{\rm G} &=&\frac 1{\pi^2} \int d\lambda_1 d\lambda_1^*d\lambda_2 d\lambda_2^* |\lambda_1-\lambda_2|^2
\prod_{k=1}^2 \frac{ e^{-\nb\lambda_k^* \lambda_k}}{(\lambda_k-u) (\lambda_k^*+\mu)}
\nn \\
&=&\frac 2{\pi^2} \int d\lambda_1 d\lambda_1^* 
(\lambda_1^*\lambda_1 - \lambda_1\lambda_2^*)
\prod_{k=1}^2 \frac{ e^{-\nb\lambda_k^* \lambda_k}}{(\lambda_k-\mu) (\lambda_k^*+\mu)} 
\nn \\
&=&\frac 2{\pi^2} \int d\lambda_1 d\lambda_1^* 
 \frac{ \lambda_1^*\lambda_1e^{-\nb\lambda_1^* \lambda_1}}{(\lambda_1-\mu) (\lambda_1^*+\mu)}
\int d\lambda_2 d\lambda_2^* 
 \frac{ e^{-\nb\lambda_2^* \lambda_2}}{(\lambda_2-\mu) (\lambda_2^*+\mu)}
\nn \\
&& + 2 \left ( \frac 1\pi\int d\lambda_1 d\lambda_1^* 
 \frac{  \lambda_1
e^{-\nb\lambda_1^* \lambda_1}}{(\lambda_1-\mu)(\lambda_1^*+\mu)} \right )^2
\nn \\
&=& 8 \int d\lambda \lambda 
 \frac{ \lambda^2 e^{-\nb\lambda^2}}{\lambda^2+\mu^2} {\rm sign}(\lambda -\mu) 
\int d\lambda \lambda 
 \frac{ e^{-\nb\lambda^2 }}{\lambda^2+\mu^2}{\rm sign}(\lambda -\mu)
\nn \\
&& + 2 \left ( 2 \int_1^\infty d\lambda \frac {\lambda \mu
e^{-\nb\lambda^2 }}{\lambda^2+\mu^2}+2 \int_0^1 d\lambda \frac {\lambda^3 
e^{-\nb\lambda^2 }}{\mu(\lambda^2+\mu^2) } \right )^2.
\ee
Using the same steps as for $n=2$,
for $n=3$ the partition function can be expressed in terms of three
 integrals
\be
Z_{3}(\mu) Z^G_{3}=
  6 Z^a_0(\mu)Z^a_1(\mu)Z^a_2(\mu)
  + 6Z^a_0(\mu)(Z^b_1(\mu))^2
          -12 Z^c_1(\mu)Z^b_1(\mu)Z^b_2(\mu)
          \ee
where
       \be
       Z_p^a(\mu)&=& \mu^{2(p+1)}\int_0^\infty dx {\rm sign}(x-1)
         \frac {e^{-\nb\mu^2x} x^p}{x+1}, \nn\\
  Z_p^b(\mu)&=& \mu^p\int_1^\infty dx  \frac {e^{-\nb\mu^2x} }{x+1}
  -(-u)^p\int_0^1 dx  \frac {e^{-\nb\mu^2x} x^p}{x+1}, \nn\\
 Z_p^c(\mu)&=& \mu^{2p+1}\int_1^\infty dx  \frac {e^{-\nb\mu^2x x^p} }{x+1}
  +\mu^{2p+1}\int_0^1 dx  \frac {e^{-\nb\mu^2x} x^{p+1}}{x+1}, \nn\\
\ee
The $n=2$ partition function can be rewritten in terms of the first two integrals
\be
Z_{2}(\mu) Z^G_{2}=
  (2 Z^a_0(\mu)Z^a_1(\mu)+ 2(Z^b_1(\mu))^2.
          \ee

\end{document}